\documentclass[12pt]{iopart}
\usepackage{iopams,ulem,color,hyperref,graphicx,url}  
\begin{document}
\title[Spontaneous collective synchronization in the Kuramoto
model with additional non-local interactions]{Spontaneous collective
synchronization in the Kuramoto
model with additional non-local interactions}
\author{Shamik Gupta}
\address{Department of Physics, Ramakrishna Mission Vivekananda
University, Belur Math, Howrah 711 202, West Bengal, India}
\ead{shamik.gupta@rkmvu.ac.in}
\vspace{10pt}
\begin{indented}
\item[]{\bf Invited contribution to the Journal of Physics A:
Math. Theor. 
{\it Special Issue}
\href{http://iopscience.iop.org/journal/1751-8121/page/Emerging-Talents}{``Emerging
Talents"} being published on the occasion of the 50th anniversary
celebrations in 2017 of the Journal of Physics series}
\end{indented}
\begin{abstract}
In the context of the celebrated Kuramoto model of globally-coupled
phase oscillators of distributed natural frequencies, which serves as a paradigm to
investigate spontaneous collective synchronization in many-body interacting systems, we report on a very rich phase diagram in
presence of thermal noise and an additional non-local interaction on a one-dimensional
periodic lattice. Remarkably, the phase
diagram involves both equilibrium and non-equilibrium phase transitions.
In two contrasting limits of the dynamics, we obtain exact analytical
results for the phase transitions. These two limits correspond to (i)
the absence of thermal noise, 
when the dynamics reduces to that of a non-linear {\it dynamical} system, and
(ii) the oscillators having the same natural frequency, when the
dynamics becomes that of a {\it statistical} system in contact with a heat bath
and relaxing to a statistical equilibrium state. In the former case, our
exact analysis is based on the use of the so-called
Ott-Antonsen ansatz to derive a reduced set of nonlinear partial
differential equations for the macroscopic evolution of the system. Our results for
the case of statistical equilibrium are on the other hand obtained by extending the
well-known transfer matrix approach for nearest-neighbor Ising model to consider non-local
interactions. The work offers a case study of exact analysis in
many-body interacting systems. The results obtained underline the crucial
role of additional non-local interactions in either destroying or
enhancing the possibility of observing synchrony in mean-field systems
exhibiting spontaneous synchronization.
\end{abstract}
\vspace{2pc}
\tableofcontents
\section{Introduction}
\label{sec:introduction}

Dynamical systems comprising a large number of interacting
constituents with a non-linear evolution in time generically exhibit
a variety of rich emergent behaviors that go well beyond the behaviour
of the constituent elements \cite{Strogatz:2014,Ott:2002}.
Perhaps the most fascinating one is that of collective synchronization, in which a large population of oscillating
units that have diverse frequencies and are interacting weakly with one another
adjust their individual rhythms to spontaneously evolve to a state in
which the units operate in unison \cite{Pikovsky:2001,Strogatz:2003}. Spontaneous synchronization lies at
the heart of many physical phenomena in nature. Even sustenance of life
requires the heart to beat as a result of harmonized contractions of the cardiac cells
in a synchronous wave. A highly non-linear cooperative
effect, synchronization is observed in yeast cell suspensions \cite{Richard:1996}, among flashing fireflies \cite{Buck:1968, Ermentrout:1991-0}, in crickets chirping in synchrony \cite{Walker:1969}, in an
audience clapping in unison for a laudable performance in a concert hall \cite{Neda:2000,Xenides:2008}, among pedestrians on footbridges
\cite{Dallard:2001}, and in a variety of experiments involving electrochemical \cite{Kiss:2002} and electronic \cite{Temirbayev:2012}
oscillators, metronomes \cite{Martens:2013}, Josephson junctions \cite{Benz:1991}, laser arrays \cite{Hirosawa:2013},
etc. The reader may refer to Ref. \cite{Strogatz:1993} for an engrossing exposition of synchronization in biological systems.

An early theoretical approach to synchrony is due to
Winfree, who couched the problem in the framework of $N \gg 1$ oscillators of nearly-identical frequencies that are weakly
coupled to one another. The weak coupling
leads to a fast relaxation of the
oscillators to their limit cycles \cite{Strogatz:2014,Ott:2002}, so that for subsequent times, they
may be characterized solely by their phases. In the
following, the word ``phase" would be used to also refer to a
thermodynamic phase of a macroscopic system. In order to avoid any possible
confusion between the two different usages of the word ``phase", we will
from now on use the term ``angle" to mean oscillator phase, and the
term ``phase" to exclusively mean a thermodynamic phase. On timescales longer than the one over which the oscillators are characterized by their angles
alone, the latter would evolve in time due to the coupling and
the frequency differences between the oscillators. To avoid
complications arising from a spatial distribution of the oscillators, Winfree endowed the system with a
mean-field geometry: every oscillator responds to the collective effect
of the whole population. The angles $\theta_j \in
[0,2\pi);~j=1,2,\ldots,N$ evolve in time as
\cite{Winfree:1967,Winfree:1980}
\begin{equation}
\frac{{\rm d}\theta_j}{{\rm d}t}=\omega_j+\frac{J}{N}\left(\sum_{k=1}^N
X(\theta_k)\right)Z(\theta_j),
\label{eq:eom-winfree}
\end{equation}
where $\omega_j$ is the natural oscillation frequency of the $j$-th
oscillator. The natural frequencies of all the oscillators
form a set of quenched disordered random
variables with a common probability distribution $G(\omega)$. In Eq.
(\ref{eq:eom-winfree}), $X(\theta_k)$ denotes the influence of the
$k$-th oscillator on the $j$-th one, which responds to the collective influence $\left(\sum_{k=1}^N
X(\theta_k)\right)$ through the sensitivity function or the so-called phase response curve $Z(\theta_j)$. The
model (\ref{eq:eom-winfree}) assumes that the functions $Z$
and $X$ are identical for all the oscillators, and that the coupling
$J>0$
is the same for every pair of oscillators. The scaling of $J$ by $N$ in
Eq. (\ref{eq:eom-winfree}) ensures that the model is well-behaved in the thermodynamic limit $N \to
\infty$. Winfree showed in simulations that for a given $J$ and for
sufficiently small diversity in the natural frequencies, the system
(\ref{eq:eom-winfree}) exhibits a synchronized state at long times. However, analytical results had 
to wait for Kuramoto, who came up with a simplification of the Winfree
model that made it amenable to an exact treatment in the thermodynamic
limit $N \to \infty$ (see however Ref. \cite{Pazo:2014} and references therein). He assumed
the weak-coupling condition $|\omega_j| \gg
J~\forall~j$, so that Eq. (\ref{eq:eom-winfree}) for every oscillator
may be averaged over its oscillation frequency, thus obtaining for
$Z(\theta_j)X(\theta_k)$ a function solely of the angle difference given
by $d(\theta_j-\theta_k)$ (see, e.g., Ref. \cite{Ermentrout:1991} for
details). He made a simple choice, $d(\theta)=\sin
\theta$, thus obtaining the dynamics \cite{Kuramoto:1975,Kuramoto:1984,Strogatz:2000,Acebron:2005,Gupta:2014-2,Rodrigues:2016,Gupta:2018}
\begin{equation}
\frac{{\rm d}\theta_j}{{\rm d}t}=\omega_j+\frac{J}{N}\sum_{k=1}^N
\sin(\theta_k-\theta_j).
\label{eq:eom-kuramoto}
\end{equation}
We show later that Eq. (\ref{eq:eom-kuramoto}) corresponds to the driven
overdamped dynamics of globally-coupled $XY$ spins. However, we
emphasize that there is no {\it a priori} justification to treat interacting limit-cycle oscillators as $XY$ spins. We note in passing that just as the Kuramoto model is obtained as
the weak-coupling limit of the Winfree model, one may also obtain (see, e.g., Refs. \cite{Golomb:2001,Politi:2015}) a
Winfree-type ensemble of oscillators by considering a suitable weak-coupling limit of the dynamics of the
so-called pulse-coupled leaky integrate-and-fire neuron system
\cite{Abbott:1993} that is extensively
employed in the field of computational neuroscience to study neuronal
dynamics.

Let us consider for $G(\omega)$ a unimodal distribution, i.e., one which is
symmetric about the mean $\Omega_0$ and decreases monotonically to
zero with increasing $|\omega-\Omega_0|$. The effect of $\Omega_0$ can be
gotten rid of from Eq. (\ref{eq:eom-kuramoto}) by viewing the dynamics in a frame
rotating uniformly with frequency $\Omega_0$ with respect to an inertial
frame; this tantamounts to implementing the Galilean shift $\theta_j \to
\theta_j+\Omega_0 t~\forall~j$ that leaves Eq. (\ref{eq:eom-kuramoto}) invariant. Denoting the
half-width-at-half-maximum (HWHM) of $G(\omega)$
by $\Delta>0$ \footnote{For a Gaussian distribution with standard
deviation equal to $\sigma_{\rm Gaussian}$, the HWHM is given by
$\Delta=\sigma_{\rm Gaussian}\sqrt{2\ln 2}$.}, we may put in evidence the dependence of the dynamics
(\ref{eq:eom-kuramoto}) on $\Delta$ by replacing the term $\omega_j$ by
$\Delta~\omega_j$, and concomitantly, consider from now on the
$\omega_j$'s as dimensionless random numbers with a common
distribution $g(\omega)$ that has zero mean and unit width and the
normalization $\int_{-\infty}^\infty {\rm d}\omega~g(\omega)=1$.
We thus obtain the dynamics
\begin{equation}
\frac{{\rm d}\theta_j}{{\rm d}t}=\Delta~\omega_j+\frac{J}{N}\sum_{k=1}^N
\sin(\theta_k-\theta_j).
\label{eq:eom-kuramoto-again}
\end{equation}

From Eq.
(\ref{eq:eom-kuramoto-again}), we note that for a given 
$g(\omega)$, the frequency term alone induces independent oscillations of every
oscillator at its own natural frequency, a tendency that is opposed by
the global coupling that favors equal angles for all the oscillators, thereby promoting global synchrony. It
is convenient to visualize the angles as points moving on
a unit circle under the dynamics (\ref{eq:eom-kuramoto-again}). A
synchronized or a clustered state then corresponds to a macroscopic 
cluster of these points that is immobile in time \footnote{This is because
the dynamics (\ref{eq:eom-kuramoto-again}) refers to a frame that is rotating
uniformly with angular frequency $\Omega_0$ with respect to an inertial frame. When viewed in the latter, however, the cluster moves around the circle at angular
frequency $\Omega_0$.}, while an unsynchronized state has points randomly distributed over the circle. For
low $J$ and initial $\theta_j$'s that are all equal, the points on the circle while starting bunched
together spread out on a timescale $\sim 1/\Delta$ due to the diversity
in the $\omega_j$'s. By contrast, for sufficiently high $J$ and an initial state with small bunching, the interaction term in
(\ref{eq:eom-kuramoto-again}) grows in time by pulling in more and more
oscillators towards the bunch, thus inducing a relaxation to a synchronized state. For a given initial
condition (or an ensemble of initial conditions), whether synchrony is 
sustained at long times and the amount of it is determined by
an interplay of the interaction with the diversity in the
natural frequencies of the oscillators.

To characterize quantitatively the amount of synchrony in the system, Kuramoto introduced the (complex) synchronization
order parameter $r(t)$ defined as \cite{Kuramoto:1984}
\begin{equation}
r(t)\equiv \frac{1}{N}\sum_{k=1}^N e^{i\theta_k(t)}.
\label{eq:r-definition}
\end{equation}
Correspondingly, one has a vector in the complex-$r$ plane with $x$ and $y$
components $(r_x,r_y)\equiv \left(\frac{1}{N}\sum_{k=1}^N \cos \theta_k,\frac{1}{N}\sum_{k=1}^N
\sin \theta_k\right)$; the length $|r|\equiv
\sqrt{r_x^2+r_y^2}$ measures the amount of synchrony, while
$\tan^{-1}(r_y/r_x)$ gives the average angle.
When the oscillators are unsynchronized so that over a stretch of time
or in an ensemble of configurations at a given time, one has with equal
probabilities $e^{i\theta}$ equal to any complex number with modulus
unity, $|r|$ averages to
zero. On the other hand, $|r|$ has a non-zero average when a finite fraction of oscillators
have angle differences that are constant in
time. We will denote by 
\begin{equation}
r^{\rm st}\equiv |r(t \to \infty)|
\label{eq:rst-definition}
\end{equation}
the stationary value of the
synchronization order parameter, which may be obtained by averaging
$r_x^2$ and $r_y^2$ over the
stationary ensemble of configurations.
Based on the discussions above, we expect $r^{\rm st}$ to exhibit qualitatively different behaviors as $\Delta$ is tuned
from low (thus favoring $r^{\rm st}\ne 0$) to
high (favoring $r^{\rm st}=0$) values at a fixed $J$.
Indeed, it has been rigorously established that under such a tuning of $\Delta$, the
system (\ref{eq:eom-kuramoto-again}) in the thermodynamic limit
undergoes in the stationary state a
continuous phase transition, from a low-$\Delta$
synchronized phase ($r^{\rm st} \ne 0$) to a high-$\Delta$ incoherent phase
($r^{\rm st}=0$), at the critical threshold $\Delta_c=\pi J g(0)/2$
\cite{Kuramoto:1984,Gupta:2014-1,Gupta:2014-2}. In the thermodynamic
limit, the system is well characterized by the probability density
function $f(\theta,\omega,t)$, defined such that $f(\theta,\omega,t){\rm
d}\theta {\rm d}\omega$ gives out of all oscillators with
frequency in $[\omega,\omega+{\rm d}\omega]$ the fraction at time $t$
that have their angle in $[\theta,\theta+{\rm d}\theta]$. While the incoherent state with
$f(\theta,\omega,t)=1/(2\pi)$ is linearly neutrally stable at all
$\Delta$'s, a stable branch corresponding to a synchronized state
bifurcates continuously for $\Delta \le \Delta_c$ \cite{Strogatz:2000}.

Over the years, the Kuramoto model has served as a paradigm to study spontaneous collective
synchronization in many-body interacting systems, and has moreover
initiated a wide variety of studies criss-crossing several disciplines
and involving physicists, mathematicians, and applied scientists. For an overview of
recent progresses and perspectives on the model and its many variants,
see Ref. \cite{Pikovsky:2015}. Results from extensive
studies of the model have found numerous applications in areas ranging from
bridge engineering and social sciences to neuroscience, and have even led to the introduction
of novel theoretical concepts in nonlinear science such as the chimera
states \cite{Abrams:2004, Abrams:2008}, see Ref. \cite{Panaggio:2015}
for a recent review. 
Chimeras are broken-symmetry states occurring in identical,
symmetrically-coupled oscillator ensembles in
which synchronized and desynchronized sub-populations coexist. These
states have been observed in a variety of experimental situations
involving, e.g., chemical and mechanical
oscillators and photoelectrochemical devices \cite{Panaggio:2015}, and
also in many theoretical frameworks besides the Kuramoto setting, e.g.,
in a system of globally-coupled complex Ginzburg-Landau oscillators
\cite{Sethia:2014} and in a network of coupled-map lattices \cite{Hagerstrom:2012}.

In this work, we investigate as to how the stationary behavior of the
mean-field Kuramoto model (\ref{eq:eom-kuramoto-again}), summarized above, gets modified by the inclusion of competing interactions that are non-local in space. Specifically, in the setting of a one-dimensional periodic
lattice with sites occupied by limit-cycle oscillators, we consider in
addition to a global coupling of the form in Eq.
(\ref{eq:eom-kuramoto-again}) a non-local coupling of strength $K$
between the angles of oscillators on one site with those of $M$
nearest-neighbor oscillators to the left and to the right. Moreover, we
consider the dynamical evolution to take place in presence of a
stochastic noise, modelled as a Gaussian, white noise with strength characterized by an effective temperature
$T$. We take the coupling $K$ to be either positive or negative. In the
former case, the non-local interaction acts in conjunction with the one
due to the global coupling in inducing synchrony in the system, thereby leading
to $r^{\rm st}\ne 0$. For $K<0$, on the other hand, the non-local interaction 
competes with the global coupling and may thus destroy the possibility of
observing synchrony in the system at long times. The dynamics of our
model is characterized by three parameters, namely, the HWHM $\Delta$ of the
frequency distribution, the non-local coupling $K$, and the temperature
$T$. Interestingly, for $\Delta=0$, when all the oscillators have the
same natural frequency of oscillation, the dynamics may be reduced to
that of a Hamiltonian system in contact with a heat bath for which the
stationary state has the usual Gibbs-Boltzmann form \cite{Huang:1987} of phase-space
distribution $\sim \exp(-H/T)$, with $H$ being the underlying
Hamiltonian. For $\Delta\ne0$, however, the dynamics relaxes at long
times to a nonequilibrium stationary state \cite{Zwanzig:2001} (in
technical terms, unlike its equilibrium counterpart, the
corresponding phase-space distribution does not satisfy detailed
balance). Thus, for general non-zero values of $\Delta,K,T$, the
dynamics of our model is dissipative, noisy, and is moreover out of
equilibrium. A combination of all of these factors, together with the non-linear
nature of the dynamical equations, offers a rather rich
playground to observe interesting collective effects, while rendering at
the same time the task of pursuing an exact analytical treatment of the
system one of great difficulty.

A powerful exact method that has been recently developed to
study non-noisy dynamics of coupled
oscillator ensembles is the proposition and the implementation of the so-called Ott-Antonsen (OA) Ansatz
\cite{Ott:2008,Ott:2009}, which allows to rewrite in the
thermodynamic limit the dynamics of coupled networks of phase
oscillators in terms of a few collective variables. Specifically, in the
context of the Kuramoto model (\ref{eq:eom-kuramoto-again}) with a Lorentzian distribution of the
oscillator frequencies, the ansatz studies the evolution in phase space
by considering in the space
${\cal D}$ of all possible phase-space distributions $f(\theta,\omega,t)$ a
particular class defined on and remaining confined to a manifold
${\cal M}$ in ${\cal D}$ under the time evolution of the angles. As a
result of the choice of the particular class of $f(\theta,\omega,t)$, one obtains a single first-order ordinary differential equation for the
evolution of the synchronization order parameter $r(t)$. The power and the usefulness of
the ansatz lies in its remarkable ability to capture precisely and
quantitatively through this single equation all, and not just
some, of the order parameter attractors and bifurcations of the dynamics
(\ref{eq:eom-kuramoto-again}) (which may be obtained by performing numerical
integration of the $N$ coupled non-linear equations
(\ref{eq:eom-kuramoto-again}) for $N \gg 1$ and evaluating $r(t)$ in numerics), for a Lorentzian $g(\omega)$. The success of the approach has led to hundreds of publications in
applied mathematics and physics; A few recent ones are Refs. \cite{Omelchenko:2014,Montbrio:2015,Ujjwal:2016,Wolfram:2016,Pazo:2016,Keeffe:2016,Petkoski:2016,Banerjee:2016,Martens:2016,Laing:2016,Ott:2017,Terada:2017,Zhang:2017}.

\begin{figure}
\centering
\includegraphics[width=130mm]{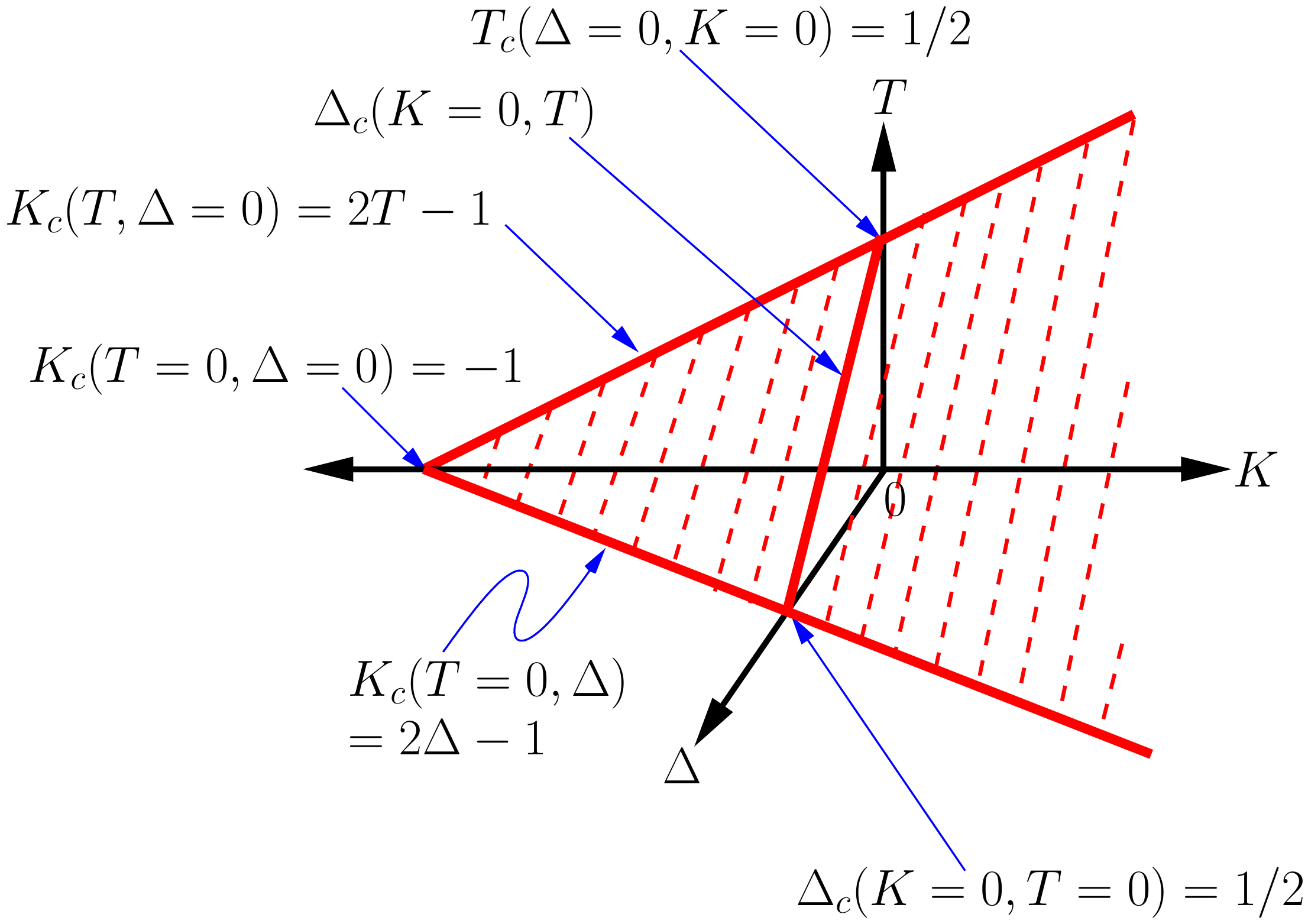}
\caption{The stationary-state phase diagram of the Kuramoto model with
additional $M$-neighbor interactions described by the dynamics
(\ref{eq:eom-general}) with a Lorentzian distribution
(\ref{eq:lorentzian-g(w)}) for the $\omega_j$'s. The phase diagram refers to the case $N \to \infty, M \to \infty, \sigma \equiv M/N < 1/2$. The thick red lines denote exact results for continuous
transition; on crossing these lines, the system undergoes
a transition between a synchronized/magnetized and an
incoherent/unmagnetized phase. Bounded by the line $K_c(T=0,\Delta)$
and the line $K_c(T,\Delta=0)$ is a surface of continuous
transition denoted schematically by dashed lines; the synchronization order parameter $r^{\rm st}$ is non-zero
inside the region bounded by the surface, and is zero outside.}
\label{fig:schematic-phase-diagram}
\end{figure}

Within the dynamical setting of our model, we
show that the system in the stationary state exhibits in the
$(\Delta,K,T)$-space a very rich phase diagram exhibiting regions of
synchronized and unsynchronized phases, and lines and surfaces of
continuous transitions between them.
The schematic phase diagram is shown in Fig.
\ref{fig:schematic-phase-diagram}. In the backdrop of the highly
non-trivial nature of the dynamics, as highlighted in previous paragraphs, exact results for the phase transitions are obtained in two contrasting limits, namely, (i) the limit $T \to
0$, and (ii) the limit $\Delta \to 0$. For case (i), the dynamics
reduces to that of a non-linear {\it dynamical} system, and our
exact analysis is based on the use of the Ott-Antonsen ansatz to derive a reduced set of nonlinear partial
differential equations for the macroscopic evolution of the system. 
On the other hand, in the case of (ii), when the
dynamics becomes that of a {\it statistical} system in contact with a heat bath
and relaxing to a statistical equilibrium state, we derive our results
by invoking the transfer matrix approach
of the nearest-neighbor Ising model well known from theories of equilibrium
statistical mechanics, and by extending it to consider non-local
interactions. Besides offering a case study of exact analysis in
many-body interacting systems, our work underlines the crucial role that non-local
interactions may play in synchronizing systems in either destroying or
enhancing the possibility of observing global synchrony in the
system.

The paper is organized as follows. In Section \ref{sec:model}, we give a
precise definition of our model and summarize known results on
stationary-state phase transitions observed in specific limits of the model. In Section \ref{sec:DeltaK-analysis}, we give
a detailed derivation of the phase diagram of the model in the
$(\Delta,K)$-plane, while the same in the $(K,T)$-plane is discussed in
Section \ref{sec:KT-analysis}. Simulation results on phase transitions
for a general point in the $(\Delta,K,T)$-space are discussed in Section
\ref{sec:DeltaKT-analysis}. The paper ends with conclusions and
perspectives in Section \ref{sec:conclusions}. A
rigorous proof that the dynamics of our model in the general case does
not verify detailed
balance 
is given in \ref{app:detailed-balance-proof}, while
\ref{app:numerics-details} contains details of a numerical scheme to
integrate the equations of motion of our model system.

\section{Model}
\label{sec:model}

Our model given by Eqs. (\ref{eq:eom-general}) and (\ref{eq:noise-final}) is a variant of the Kuramoto model (\ref{eq:eom-kuramoto-again}). To
derive it, let us first consider a one-dimensional ($1d$) lattice of $N$ sites with periodic
boundary conditions (site $j+N \equiv$ site $j$, with $j=1,2,\ldots,N$),
where each site is occupied by a limit-cycle oscillator that is
characterized completely by its angle $\theta_j \in [0,2\pi)$ and its natural
frequency $\omega_j \in [-\infty,\infty]$. As in the Kuramoto model
(\ref{eq:eom-kuramoto-again}), the $\omega_j$'s are dimensionless
numbers distributed according to a common unimodal distribution $g(\omega)$ that has zero mean and unit width, with the
normalization $\int_{-\infty}^\infty {\rm d}\omega~g(\omega)=1$. The angles evolve in time according to
the dynamics
\begin{equation}
\frac{{\rm d}\theta_j}{{\rm d}t}=\Delta~\omega_j+\frac{J}{N}\sum_{k=1}^N
\sin(\theta_k-\theta_j)+\frac{K}{2M}\sum_{k=-M}^M
\sin(\theta_{j+k}-\theta_j),
\label{eq:eom-our-model}
\end{equation}
where the second term on the right hand side is the usual global coupling of the Kuramoto
model, while the additional term that we introduce is the non-local
($M$-neighbor) interaction between the oscillators represented by the
third term on the right hand side. Here, $K$ stands
for the strength of coupling between an oscillator on a site with each of $M$ neighboring oscillators to the left and to the
right, with $K>0$ (respectively, $K<0$) implying
attractive (respectively, repulsive) interaction. Note
that for $M=N/2$, the model (\ref{eq:eom-our-model}) reduces to the Kuramoto
model with a global coupling constant equal to $J+K$. Our aim in this work is to study the modification to
the Kuramoto behavior due to non-local interactions, Hence, we consider the
allowed range of values of $M$ to be satisfying $M<N/2$. Setting $K$ to
zero allows to recover the Kuramoto model (\ref{eq:eom-kuramoto-again}).
In passing, we note that a dynamics similar to Eq.
(\ref{eq:eom-our-model}), but without the global coupling term and with a phase lag in the
interaction (that is, having the $M$-neighbor interaction to be of the form
$\sin(\theta_{j+k}-\theta_j-\alpha)$, with $\alpha \in
(0,\pi/2)$ being the phase lag) has been analyzed in Ref.
\cite{Omelchenko:2014}; as is well known, the presence of a phase lag
has important consequences on the behavior of the Kuramoto model
\cite{Sakaguchi:1986}.
Reference \cite{Laing:2016} considered a model similar to Eq.
(\ref{eq:eom-our-model}), but without the
inclusion of the global coupling term. We note that as regards observing chimera states, the essential
dynamical setup introduced, e.g., in Ref. \cite{Wolfrum:2011}, is given by the equation of motion 
\begin{equation}
\frac{{\rm d}\theta_j}{{\rm d}t}=\omega+\frac{1}{2M}\sum_{k=-M}^M
\sin(\theta_{j+k}-\theta_j-\alpha),
\label{eq:eom-chimera}
\end{equation}
with $\omega$ being the common natural frequency of the oscillators,
and with $\alpha \in (0,\pi/2)$ being the phase-lag parameter. We thus
observe an important difference in the
form of the non-local interaction in (\ref{eq:eom-chimera}) with respect to
the dynamics (\ref{eq:eom-our-model}), namely, the presence of the
phase lag $\alpha$, whose inclusion has been argued to be crucial for observing the chimeras
\cite{Panaggio:2015}. The issue of whether chimeras are observed in the
dynamics (\ref{eq:eom-our-model}) on including a phase lag in the
non-local interaction and on making all the natural frequencies to be
identical is an important question that is relegated to
future studies.

A representative example of $g(\omega)$ that we specifically
consider in this work to demonstrate our results and for which we obtain exact analytical results for relevant macroscopic properties is that of a Lorentzian distribution:
\begin{equation}
g(\omega)=\frac{1}{\pi}\frac{1}{\omega^2+1}.
\label{eq:lorentzian-g(w)}
\end{equation}

Before proceeding, we rewrite Eq. (\ref{eq:eom-our-model}) in a dimensionless form, by defining
dimensionless quantities 
\begin{equation}
\overline{t}\equiv Jt,~\overline{\Delta}\equiv \frac{\Delta}{J},
~\overline{K}\equiv \frac{K}{J};
\label{eq:dimensionless-quantities-no-noise-overdamped}
\end{equation}
we get the dimensionless equation
\begin{equation}
\frac{{\rm d}\theta_j}{{\rm d}\overline{t}}=\overline{\Delta}~\omega_j+\frac{1}{N}\sum_{k=1}^N
\sin(\theta_k-\theta_j)+\frac{\overline{K}}{2M}\sum_{k=-M}^M
\sin(\theta_{j+k}-\theta_j).
\label{eq:eom}
\end{equation}
%

\subsection{Relation to long-range interacting systems}
\label{sec:LRI}
We now establish a relation of the dynamics (\ref{eq:eom}) to a specific
limit of a certain Hamiltonian dynamics, which would prove quite useful
later in the paper in studying the model. To this end, let us consider the Hamiltonian of a
mean-field (classical) XY model in presence of additional non-local interactions on a
$1d$ periodic lattice:
\begin{eqnarray}
\fl H=\sum_{j=1}^N\frac{p_j^2}{2m}+\frac{\widetilde{J}}{2N}\sum_{j,k=1}^N\left[1-\cos(\theta_j-\theta_k)\right]
-\frac{\widetilde{K}}{4M}\sum_{j=1}^N\sum_{k=-M}^M\cos(\theta_{j+k}-\theta_j),
\label{eq:H} 
\end{eqnarray}
where $p_j\equiv mv_j$ is the momentum conjugate to $\theta_j$ ($p_j$
and $\theta_j$ together constitute the set of canonically conjugate dynamical variables associated with
the site $j$), $m$ is
the mass, $v_j$ is the velocity, while $\widetilde{J}>0$ and
$\widetilde{K}$ are respectively the global and the non-local
($M$-neighbor) coupling
constant. 

Note that in the Hamiltonian (\ref{eq:H}), the global-coupling term
involves every spin interacting with every other with the same strength
$\widetilde{J}$. Such an interaction is the extreme form (the mean-field
limit) of the so-called
long-range interactions exhibited by physical systems. 
Long-range interacting (LRI) systems are those in which the constituent
particles interact with each other with a strength that decays slowly
with their separation $r$ as
$r^{-\alpha}$ for large $r$, with $0 \le \alpha \le d$ in $d$ spatial
dimensions \cite{Campa:2009,Bouchet:2010,Campa:2014,Levin:2014,Gupta:2017}.
LRI systems are encountered across disciplines, in astrophysics,
hydrodynamics, plasmas, atomic and nuclear physics, and condensed matter
physics. These systems are intrinsically non-additive so that they cannot be
trivially divided into independent macroscopic sub-parts, a feature that leads to many fascinating
phenomena not observed with short-range interactions, e.g.,
inequivalence of statistical ensembles \cite{Bouchet:2010,Campa:2014}. Other striking effects are breaking of ergodicity: the phase space is broken up into subspaces not connected by local dynamics. A very interesting dynamical feature of LRI systems is the occurrence of quasistationary
states during relaxation to equilibrium. These states involve a slow relaxation of macroscopic
observables over times that diverge algebraically with the system size, so that in the thermodynamic limit,
the system remains trapped in them and never attains the Boltzmann-Gibbs
equilibrium \cite{Yamaguchi:2004,Levin:2014,Campa:2014}.

The Hamiltonian (\ref{eq:H}) has in addition to a long-range interaction
a short-range one described by the coupling among $M$ nearest-neighbors.
For small $M$, we expect the long-range behavior to dominate, and
indeed, the equilibrium properties of the Hamiltonian (\ref{eq:H})
with
$M=1$ and within microcanonical and canonical ensembles have demonstrated the feature of
ensemble inequivalence emerging as a consequence of long-range
interactions \cite{Campa:2006,Dauxois:2010}. 

In contact with a 
heat bath that induces noise into the system, and in presence of a
friction constant $\gamma>0$, the dynamics derived from the Hamiltonian
(\ref{eq:H}) and with additional external drives in the form of quenched
disordered external toques $\widetilde{\Delta}~\omega_j$ acting on the individual spins is given by the set of
equations
\begin{eqnarray}
\fl \frac{{\rm d}\theta_j}{{\rm d}t}=v_j, \nonumber \\
\label{eq:H-eom} \\
\fl m\frac{{\rm d}v_j}{{\rm d}t}=\gamma \widetilde{\Delta}~\omega_j-\gamma
v_j+\frac{\widetilde{J}}{N}\sum_{k=1}^N
\sin(\theta_k-\theta_j)+\frac{\widetilde{K}}{2M}\sum_{k=-M}^M
\sin(\theta_{j+k}-\theta_j)+\sqrt{\gamma}~\eta_j(t), \nonumber 
\end{eqnarray}
where $\widetilde{\Delta}>0$ is a given parameter characterizing
the strength of the external torques, while $\eta_j(t)$ is a Gaussian, white noise with 
\begin{equation}
\langle \eta_j(t)\rangle=0,~\langle \eta_j(t)\eta_k(t')
\rangle=2T\delta_{jk}\delta(t-t').
\label{eq:noise-properties}
\end{equation}
Here, $T$ is the temperature of the heat bath in units of the Boltzmann
constant, while  
angular brackets denote averaging over noise realizations. 
Let us now define the following dimensionless
quantities:
\begin{equation}
\overline{t}\equiv
t\frac{\widetilde{J}}{\gamma},~\overline{\Delta}\equiv \gamma
\frac{\widetilde{\Delta}}{\widetilde{J}},~\overline{\eta}_i(\overline{t})\equiv\eta_i(t)\sqrt{\frac{\gamma}{\widetilde{J}}},~\overline{K}\equiv
\frac{\widetilde{K}}{\widetilde{J}},~\overline{T}\equiv
\frac{T}{\widetilde{J}},
\label{eq:dimensionless-quantities-overdamped-noise}
\end{equation}
where note that for given values of $\widetilde{J}$ and $\widetilde{K}$, the ratio
$\widetilde{K}/\widetilde{J}$ may not equal the quantity $\overline{K}$
as defined in the paragraph preceding Eq. (\ref{eq:eom}); if this is the case, the equality may be achieved by multiplying both
$\widetilde{J}$ and $\widetilde{K}$ by the same factor. Similarly, for
given values of $\gamma, \widetilde{J}$ and $\widetilde{\Delta}$, the ratio
$\gamma \widetilde{\Delta}/\widetilde{J}$ may not equal the quantity
$\overline{\Delta}$
as defined in the paragraph preceding Eq. (\ref{eq:eom}), and when this is the case, the equality may be achieved by multiplying both
$\widetilde{J}$ and $\widetilde{\Delta}$ by the same factor. Using the
definitions in Eq. (\ref{eq:dimensionless-quantities-overdamped-noise}), we obtain from Eq.
(\ref{eq:H-eom}) and in the limit $m/\gamma\ll 1$ the overdamped dynamics
\begin{eqnarray}
&&\frac{{\rm d}\theta_j}{{\rm
d}\overline{t}}=\overline{\Delta}~\omega_j+\frac{1}{N}\sum_{k=1}^N
\sin(\theta_k-\theta_j)+\frac{\overline{K}}{2M}\sum_{k=-M}^M
\sin(\theta_{j+k}-\theta_j)+\overline{\eta}_j(\overline{t}), 
\label{eq:H-eom-overdamped-dimensionless} 
\end{eqnarray}
with
\begin{equation}
\langle \overline{\eta}_j(\overline{t})\rangle=0,~\langle
\overline{\eta}_j(\overline{t})\overline{\eta}_k(\overline{t}')
\rangle=2\overline{T}\delta_{jk}\delta(\overline{t}-\overline{t}').
\label{eq:noise-properties-dimensionless}
\end{equation}
From Eqs. (\ref{eq:H-eom-overdamped-dimensionless}) and
(\ref{eq:noise-properties-dimensionless}), it is evident that
as $\overline{T} \to 0$, the overdamped dynamics (\ref{eq:H-eom-overdamped-dimensionless}) reduces to Eq. (\ref{eq:eom}). 

On the basis of the foregoing discussions, we conclude that the general first-order
dynamics that incorporates in specific limit the dynamics
(\ref{eq:eom}) is given by
\begin{eqnarray}
&&\frac{{\rm d}\theta_j}{{\rm d}t}=\Delta~\omega_j+\frac{1}{N}\sum_{k=1}^N
\sin(\theta_k-\theta_j)+\frac{K}{2M}\sum_{k=-M}^M
\sin(\theta_{j+k}-\theta_j)+\eta_j(t); \label{eq:eom-general} \\
&&\langle \eta_j(t)\rangle=0,~\langle
\eta_j(t)\eta_k(t')
\rangle=2T\delta_{jk}\delta(t-t'), \label{eq:noise-final}
\end{eqnarray}
where all the quantities are dimensionless, and we have dropped the
overbars in order not to overload our notation. Equations (\ref{eq:eom-general})
and (\ref{eq:noise-final}) define our model of interest in this work.

\subsection{Summary of known results and our queries}
\label{sec:phase-transition}

We may ask: what is the nature of the stationary state that the dynamics
(\ref{eq:eom-general}) relaxes to at long times (i.e., in the limit $t
\to \infty$)? For $\Delta=0$, the dynamics (\ref{eq:eom-general}) relaxes to a 
Boltzmann-Gibbs (BG) equilibrium state, so that the probability distribution
of the angles $\{\theta_j\}_{1\le j \le N}$ has the usual form
\begin{equation}
P_{\rm eq}(\{\theta_j\})\propto \exp[-{\cal V}(\{\theta_j\})/T],
\label{eq:Delta0-equilibrium-distribution}
\end{equation}
with ${\cal V}$ being a potential energy function:
\begin{equation}
{\cal V}(\{\theta_j\})\equiv
\frac{1}{2N}\sum_{j,k=1}^N\left[1-\cos(\theta_j-\theta_k)\right]-\frac{K}{4M}\sum_{j=1}^N\sum_{k=-M}^M
\cos(\theta_{j+k}-\theta_j);
\label{eq:V}
\end{equation}
for a proof, see \ref{app:detailed-balance-proof}. For $\Delta \ne 0$, the dynamics
(\ref{eq:eom-general}) does not correspond to
a Hamiltonian system because of the natural frequency term that
cannot be derived from any potential $V_{\rm pot}(\{\theta_j\})$ that
satisfies the periodicity of
the system, namely, $V_{\rm pot}(\{\theta_j+2\pi\})=
V_{\rm pot}(\{\theta_j\})$, and thus be intrinsic to the system. The
external drives in the form of the natural frequencies continuously pump energy into the system. In this case, the
dynamics at long times relaxes to a nonequilibrium stationary state
\cite{Zwanzig:2001}, which does not have the BG form of angle distribution, and
which violates detailed balance. The latter property is
proven in \ref{app:detailed-balance-proof}.

Note that the dynamics (\ref{eq:eom-general}) is characterized by three
dimensionless parameters $(\Delta,K,T)$, see Fig.
\ref{fig:schematic-phase-diagram}. In this work, we consider the dynamics in the
thermodynamic limit, and obtain its stationary-state phase diagram in
the $(\Delta,K,T)$-space. It is pertinent to discuss
the range of values of $M$ our results for the phase diagram apply to. To this
end, let us define an interaction radius $\sigma$ as $\sigma\equiv M/N$.
Note that our model (\ref{eq:eom-general}) is to be considered for
$M<N/2$, that is, for $\sigma <1/2$. Suppose one takes
first the thermodynamic limit $N \to \infty$, and then gradually
increase $M$ to larger and larger values while keeping $\sigma<1/2$.
The limiting phase diagram that one gets as $M$ approaches infinity is 
what we obtain in this work, with the corresponding analysis for the $(\Delta,K)$-plane described in Section \ref{sec:DeltaK-analysis}
and that for the $(K,T)$-plane described in Section
\ref{sec:KT-analysis}. We will see that the actual value of $\sigma$ has
only the role of a parameter that characterizes the dynamics of
the system.

We note that, to the best of our knowledge, our quest for the complete phase diagram of the model (\ref{eq:eom-general}) has not
been addressed before. Only certain
limits of the dynamics and the associated phase diagrams have been
considered by two different communities of physicists, namely, the
dynamical physicists and the statistical physicists. We now summarize these contributions.

Before proceeding, let us discuss
qualitatively some
general features of the dynamics. Considering Eq.
(\ref{eq:eom-general}) and a given frequency distribution
$g(\omega)$, we note that the effect of the frequency term and the noise
term is to induce every oscillator to oscillate at its natural frequency
on an average, with thermal fluctuations superimposed on the average
behavior.  In contrast to the Kuramoto model (\ref{eq:eom-kuramoto-again}), this
tendency is now opposed by both the global and the non-local coupling
among the oscillators. While the former favors equal angles for the oscillators, thereby promoting
{\it global} synchrony among all the oscillators, the latter induces a
{\it local} (that is, among $M$ neighboring oscillators to the left and
to the right of a given oscillator) order. The latter can be either ferromagnetic or antiferromagnetic (borrowing terminologies from spin systems), depending respectively on whether $K$ is positive or negative.
The interplay of these various tendencies ultimately determines whether synchrony among the
oscillators is
sustained in the stationary state and the amount of it.
In this backdrop, we now summarize the known phase transitions exhibited by the
dynamics (\ref{eq:eom-general}).

\begin{itemize}
\item The case $K=T=0$ corresponds to the Kuramoto
model, which is thus confined to the
$\Delta$-axis, see Fig. \ref{fig:schematic-phase-diagram}. As already discussed, the system in the stationary state undergoes a
continuous phase transition as a function of $\Delta$, from a low-$\Delta$
synchronized phase to a high-$\Delta$ incoherent phase at the critical threshold $\Delta_c(K=0,T=0)=\pi g(0)/2$
\cite{Kuramoto:1984,Gupta:2014-1,Gupta:2014-2}. For the Lorentzian
distribution (\ref{eq:lorentzian-g(w)}), we find that $\Delta_c(K=0,T=0)=1/2$.
\item The case $K=\Delta=0$ corresponds to the so-called Brownian
mean-field (BMF) model \cite{Chavanis:2014}, a set-up to study statics and dynamics of LRI systems in contact with an external heat
bath. The underlying Hamiltonian is
obtained from Eq. (\ref{eq:H}) by setting $\widetilde{K}$ to zero. The
Hamiltonian describes a system of globally-coupled (classical) $XY$ spins, which
allows to draw analogies with magnetic systems, and to refer to the
corresponding stationary phases, the synchronized and the incoherent
phase, as the magnetized and the
unmagnetized phase, respectively. In equilibrium, the system exhibits a continuous
transition between the two phases at the critical temperature
$T_c(\Delta=0,K=0)=1/2$.
\item The case $K=0, \Delta \ne 0, T \ne 0$ corresponds to the
Kuramoto dynamics in presence of Gaussian, white noise, which was
studied to account for stochastic fluctuations of the $\omega_j$'s
in time \cite{Sakaguchi:1988}. In the stationary state, the transition
point $\Delta_c(K=0,T=0)$, mentioned above,
goes over to become a line of continuous transition between the
synchronized and the incoherent phase, whose equation
$\Delta_c=\Delta_c(K=0,T)$ is obtained by solving
\cite{Sakaguchi:1988,Gupta:2014-1}
\begin{equation}
2=\int_{-\infty}^\infty {\rm
d}\omega~\frac{Tg(\omega)}{T^2+\omega^2\Delta_c^2(K=0,T)}.
\label{eq:Sakaguchi-transition-line}
\end{equation}
For the Lorentzian distribution (\ref{eq:lorentzian-g(w)}), one may
evaluate the integral on the right hand side of Eq.
(\ref{eq:Sakaguchi-transition-line}) by converting
it to a complex integral, and then choosing a contour consisting of the
real-$\omega$ axis closed on the lower-half
complex-$\omega$ plane by an infinite semicircle on which the integral gives zero contribution. Evaluating the integral and using Eq.
(\ref{eq:Sakaguchi-transition-line}), one obtains
\begin{equation}
\Delta_c(K=0,T)=\frac{1}{2}-T.
\label{eq:phase-transition-Delta-T-Lorentzian}
\end{equation}
From Eq. (\ref{eq:phase-transition-Delta-T-Lorentzian}), it is easily checked that the line
$\Delta_c=\Delta_c(K=0,T)$ has an
intercept on the $T$-axis equal to $1/2$, in agreement with the
phase transition point for the BMF model mentioned above.
\end{itemize}
The aforementioned points and lines of continuous transitions, with
$r^{\rm st}$ as the order parameter, are
indicated schematically in the $(\Delta,T)$-plane in Fig.
\ref{fig:schematic-phase-diagram}. Note that these transitions all refer
to the mean-field limit of the dynamics (\ref{eq:eom-general}). In 
this paper, our primary objective is to investigate as to how this mean-field
behavior is modified by the inclusion of the
$M$-neighbor interaction. In other words, referring to Fig.
\ref{fig:schematic-phase-diagram}, we ask: how does the phase diagram
in the $(\Delta,T)$-plane extend to the whole of the
$(\Delta,K,T)$-space ? 
In the rest of the paper, we use interchangeably the terms ``magnetized"
and ``synchronized" to describe the
clustered phase, and the terms ``unmagnetized" and ``homogeneous" for the unsynchronized/incoherent phase in model (\ref{eq:eom-general}).

\section{Analysis for the $(\Delta,K)$-plane with $T=0$}
\label{sec:DeltaK-analysis}

In this section, we consider the dynamics (\ref{eq:eom-general}) with $T=0$, and turn to
a discussion of its stationary state properties. To this end, let us
introduce in the spirit of $r(t)$ a local synchronization order
parameter $Z_j(t)$ defined at the $j$-th site by the equation \cite{Omelchenko:2014}
\begin{equation}
Z_j(t)\equiv \frac{1}{2M}\sum_{k=-M}^M e^{i\theta_{j+k}(t)}.
\label{eq:Zj-definition}
\end{equation}
In terms of $Z_j(t)$ and $r(t)$, the equation of motion
(\ref{eq:eom-general}) with $T=0$ takes a form convenient for
further analysis in the present section:
\begin{equation}
\frac{{\rm d}\theta_j}{{\rm
d}t}=\Delta~\omega_j+\frac{1}{2i}[r(t)e^{-i\theta_j}-r^\star(t)
e^{i\theta_j}]+\frac{K}{2i}[Z_j(t) e^{-i\theta_j}-Z_j^\star(t)
e^{i\theta_j}],
\label{eq:eom-Zj-r}
\end{equation}
where $\star$ denotes complex conjugation. From Eq. (\ref{eq:eom-Zj-r}),
it is evident that $r(t)$ (respectively, $Z_j(t)$) plays the role of a
complex global (respectively, local) mean field, with both
driving the dynamical evolution of the angles. 

As mentioned in Section \ref{sec:model}, in order to obtain our desired phase diagram, we consider the limits $N \to \infty$ and $M
\to \infty$, keeping $\sigma=M/N<1/2$. In such a situation, it is
reasonable and convenient to invoke a continuum limit of the dynamics in
order to pursue its analytical treatment.
The continuum limit corresponds to fixing the total length of the
periodic lattice to be $2\pi$ and denoting the spatial location of the
$j$-th site by $x_j \equiv 2\pi j/N$, so that
as $N \to \infty$, the
variable $x_j$ turns into a continuous variable $x \in [0,2\pi)$. In the
continuum limit, the system is characterized by
the probability density function $f(\theta,\omega,x,t)$, defined such
that $f(\theta,\omega,x,t){\rm d}\theta {\rm d}\omega {\rm d}x$ gives
the probability at time $t$ that an oscillator in position $[x,x+{\rm d}x]$ and with
its natural frequency in $[\omega,\omega+{\rm d}\omega]$ has its angle
in $[\theta,\theta+{\rm d}\theta]$. The density function satisfies
$f(\theta+2\pi,\omega,x,t)=f(\theta,\omega,x,t)$, and the normalization
\begin{equation}
\int_0^{2\pi}{\rm
d}\theta~f(\theta,\omega,x,t)=g(\omega)~\forall~x,t.
\label{eq:normalization-f}
\end{equation}
In the continuum limit, the local
mean field becomes
\begin{equation}
Z(x,t)=\int_0^{2\pi}{\rm d}y~G(x-y)\int_{-\infty}^\infty {\rm
d}\omega\int_0^{2\pi}{\rm d}\theta~e^{i\theta}f(\theta,\omega,y,t),
\label{eq:Zj-definition-continuum}
\end{equation}
with
\begin{equation}
G(x)=\left\{\begin{array}{ll}
\frac{1}{4\pi \sigma} ~{\rm if}~|x|<2\pi \sigma, \\ 
0 ~{\rm otherwise}, \\ 
\end{array}
\right. \\
\label{eq:G-definition}
\end{equation}
while the Kuramoto order parameter becomes
\begin{equation}
r(t)=\frac{1}{2\pi}\int_0^{2\pi}{\rm d}x~\int_{-\infty}^\infty {\rm
d}\omega\int_0^{2\pi}{\rm d}\theta~e^{i\theta}f(\theta,\omega,x,t).
\label{eq:r-definition-continuum}
\end{equation}
In view of having a periodic spatial domain, all spatial
integrals are to be evaluated using periodic boundary conditions. Note
that one has $\int_0^{2\pi}{\rm d}x~Z(x,t)=2\pi~r(t)$.

The density $f(\theta,\omega,x,t)$ evolves in time according to the
continuity equation that follows from the conservation of the total
number of oscillators under the dynamics (\ref{eq:eom-Zj-r}):
\begin{eqnarray}
\fl \frac{\partial f}{\partial t}+\frac{\partial }{\partial
\theta}\Big[\Big(\Delta~\omega+\frac{1}{2i}[r(t)e^{-i\theta}-r^\star(t)
e^{i\theta}]+\frac{K}{2i}[Z(x,t)e^{-i\theta}-Z^\star(x,t)
e^{i\theta}]\Big)f\Big]=0.
\label{eq:continuity-equation}
\end{eqnarray}
Being $2\pi$-periodic in $\theta$, we expand the density
$f(\theta,\omega,x,t)$ in a Fourier series in $\theta$:
\begin{equation}
f(\theta,\omega,x,t)=\frac{g(\omega)}{2\pi}\left[1+\sum_{n=1}^\infty
\Big(\widetilde{f}_n(\omega,x,t)e^{in\theta}+[\widetilde{f}_n(\omega,x,t)]^\star
e^{-in\theta}\Big)\right],
\label{eq:f-expansion}
\end{equation}
where $\widetilde{f}_n(\omega,x,t)$ is the $n$-th Fourier coefficient. Using $\int_0^{2\pi}{\rm
d}\theta~e^{in\theta}=2\pi \delta_{n,0}$, we check that the above expansion
satisfies Eq. (\ref{eq:normalization-f}).

We now implement the Ott-Antonsen (OA) ansatz that consists in
restricting to the class of Fourier coefficients \cite{Ott:2008,Ott:2009}
\begin{equation}
\widetilde{f}_n(\omega,x,t)=[\alpha(\omega,x,t)]^n,
\label{eq:OA}
\end{equation}
with $\alpha(\omega,x,t)$ an arbitrary function, and with
$|\alpha(\omega,x,t)|<1$, so that the infinite series in Eq.
(\ref{eq:f-expansion}) is
converging. The OA ansatz also assumes that $\alpha(\omega,x,t)$ may be
analytically continued to the whole of the complex-$\omega$ plane, that
it has no
singularities in the lower-half complex-$\omega$ plane, and that
$|\alpha(\omega,x,t)|\to 0$ as ${\rm Im}(\omega) \to -\infty$
\cite{Ott:2008,Ott:2009}. 

Using Eqs. (\ref{eq:f-expansion}) and (\ref{eq:OA}) in Eqs.
(\ref{eq:Zj-definition-continuum}) and (\ref{eq:r-definition-continuum}), we obtain
\begin{eqnarray}
&&r(t)=\frac{1}{2\pi}\int_0^{2\pi}{\rm d}x\int_{-\infty}^\infty {\rm
d}\omega~g(\omega)\alpha^\star(\omega,x,t), \label{eq:r-OA} \\
&&Z(x,t)=\int_0^{2\pi}{\rm d}y~G(x-y)\int_{-\infty}^\infty {\rm
d}\omega~ g(\omega)\alpha^\star(\omega,y,t).
\label{eq:Zj-OA}
\end{eqnarray}
Equations (\ref{eq:f-expansion}) and (\ref{eq:OA}) and the above expressions
for $r(t)$ and $Z(x,t)$ on substituting in Eq.
(\ref{eq:continuity-equation}), and then on collecting and equating the
coefficient of $e^{in\theta}$ to zero yield 
\begin{eqnarray}
&\frac{\partial \alpha(\omega,x,t)}{\partial t}+i\Delta~\omega
\alpha(\omega,x,t)+\frac{1}{2}[r(t)\alpha^2(\omega,x,t)-r^\star(t)]\nonumber
\\
&+\frac{K}{2}[Z(x,t)\alpha^2(\omega,x,t)-Z^\star(x,t)]=0.
\label{eq:alpha-differential-equation}
\end{eqnarray}

\begin{figure}[!ht]
\centering
\includegraphics[width=90mm]{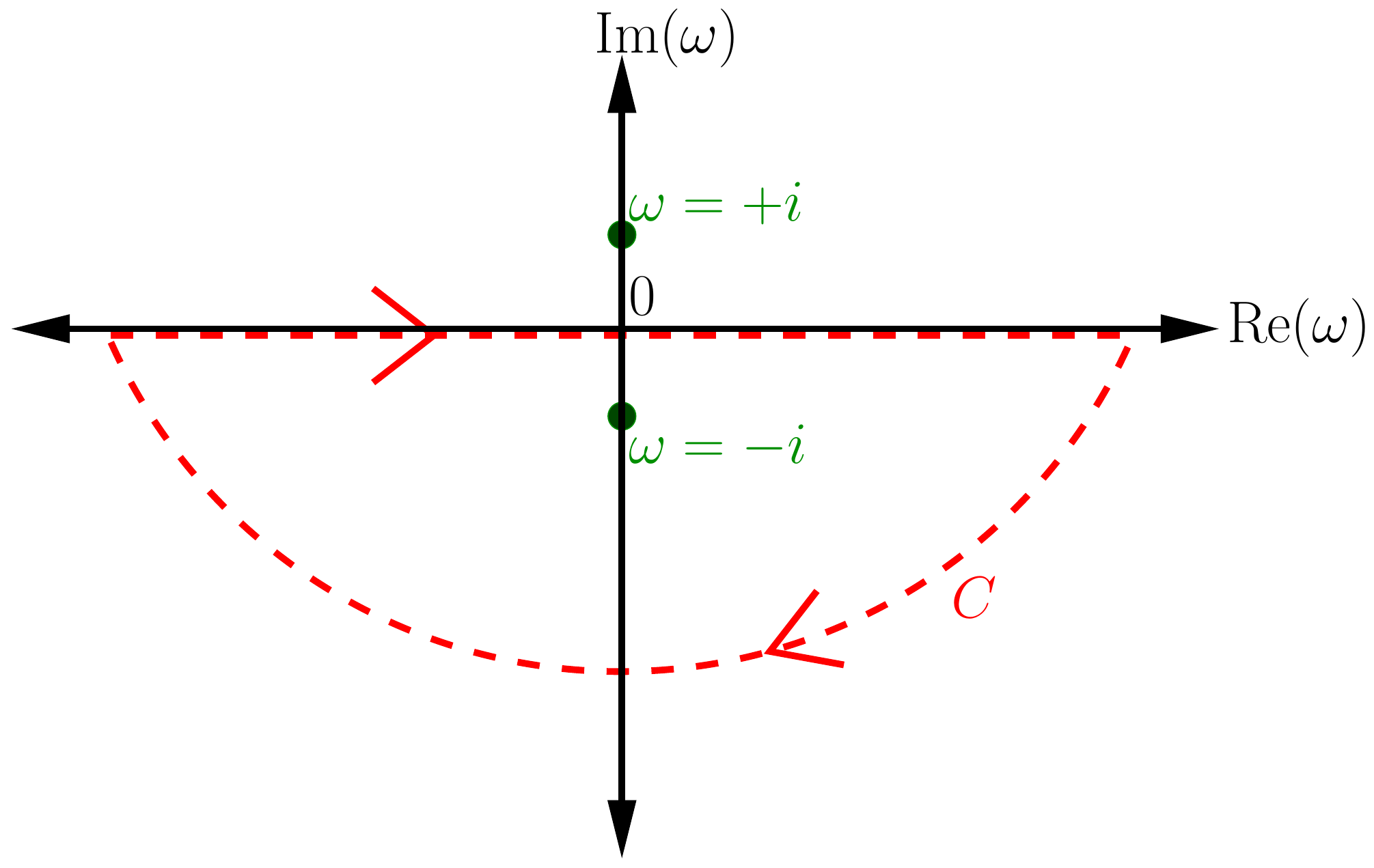}
\caption{The contour $C$ in the complex-$\omega$ plane to perform the
integration in Eqs. (\ref{eq:r-integral}) and (\ref{eq:Zj-integral}).
Also shown are the poles of the integrand at $\omega=\pm i$.}
\label{fig:contour}
\end{figure}

For the Lorentzian $g(\omega)$, Eq. (\ref{eq:lorentzian-g(w)}), one may
evaluate $r(t)$ and $Z(x,t)$ by using Eq. (\ref{eq:lorentzian-g(w)}) in 
Eqs. (\ref{eq:r-OA}) and
(\ref{eq:Zj-OA}) to get
\begin{eqnarray}
r(t)&=&\frac{1}{4i\pi^2}\int_0^{2\pi}{\rm
d}x\oint_C {\rm
d}\omega~\alpha^\star(\omega,x,t)\left[\frac{1}{\omega-i}-\frac{1}{\omega+i}\right],
\label{eq:r-integral}
\end{eqnarray}
and
\begin{eqnarray}
Z(x,t)&=&\frac{1}{2i\pi}\int_0^{2\pi}{\rm
d}y~G(x-y)\oint_C {\rm
d}\omega~\alpha^\star(\omega,y,t)\left[\frac{1}{\omega-i}-\frac{1}{\omega+i}\right],
\label{eq:Zj-integral}
\end{eqnarray}
where the contour $C$ is shown in
Fig. \ref{fig:contour}, and we have used the fact that the contribution to the integral
from the semicircular part of the contour vanishes in view of
$|\alpha(\omega,x,t)|\to 0$ as ${\rm Im}(\omega) \to -\infty$. Evaluating the integrals in Eqs.
(\ref{eq:r-integral}) and (\ref{eq:Zj-integral}) by the residue theorem, we get
\begin{eqnarray}
&&r(t)=\frac{1}{2\pi}\int_0^{2\pi}{\rm
d}x~\alpha^\star(-i,x,t), \label{eq:r-exact} \\
&&Z(x,t)=\int_0^{2\pi}{\rm
d}y~G(x-y) \alpha^\star(-i,y,t).
\label{eq:Zj-exact}
\end{eqnarray}

Calling $\alpha(-i,x,t)=u(x,t)$, Eq.
(\ref{eq:alpha-differential-equation}) then gives
\begin{eqnarray}
&&\frac{\partial u(x,t)}{\partial
t}+\Delta~u+\frac{1}{2}[r(t)u^2(x,t)-r^\star(t)]\nonumber \\
&&+\frac{K}{2}[Z(x,t)u^2(x,t)-Z^\star(x,t)]=0.
\label{eq:u-differential-equation}
\end{eqnarray}
The stationary solution $u_{\rm st}(x)$ of the above equation satisfies
\begin{equation}
\Delta~u_{\rm st}(x)+\frac{1}{2}[r_{\rm st}u_{\rm st}^2(x)-r_{\rm st}^\star]+\frac{K}{2}[Z_{\rm st}(x)u_{\rm st}^2(x)-Z_{\rm st}^\star(x)]=0,
\label{eq:u-differential-equation-stationary}
\end{equation}
with
\begin{eqnarray}
&&r_{\rm st}=\frac{1}{2\pi}\int_0^{2\pi}{\rm
d}x~u_{\rm st}^\star(x), \label{eq:r-exact-stationary} \\
&&Z_{\rm st}(x)=\int_0^{2\pi}{\rm
d}y~G(x-y) u_{\rm st}^\star(y).
\label{eq:Zj-exact-stationary}
\end{eqnarray}
%

\subsection{Uniformly incoherent state: Stability}
\label{sec:uniformly-incoherent-state-stability}

The uniformly incoherent state
$u^{\rm inc}_{\rm st}(x)=0~\forall~x$, yielding
$r_{\rm st}=0$ and $Z_{\rm st}(x)=0~\forall~x$, evidently satisfies Eq.
(\ref{eq:u-differential-equation-stationary}), and is thus a stationary
solution of Eq. (\ref{eq:u-differential-equation}). Let us study the
linear stability of such a state by linearizing Eq.
(\ref{eq:u-differential-equation}) about the state. To this end, we
write
\begin{equation}
u(x,t)=u^{\rm inc}_{\rm st}(x)+\delta u(x,t);~~|\delta u(x,t)|\ll 1,
\label{eq:u-expansion-incoherent-state}
\end{equation}
which on using in Eq. (\ref{eq:u-differential-equation}) yields to
leading order in $\delta u$ the equation
\begin{eqnarray}
&&\frac{\partial \delta u(x,t)}{\partial
t}+\Delta~\delta
u(x,t)-\frac{\delta r^\star(t)}{2}-\frac{K\delta Z^\star(x,t)}{2}=0,
\label{eq:u-differential-equation-incoherent-state-linearized}
\end{eqnarray}
with
\begin{eqnarray}
&&\delta r^\star(t)=\frac{1}{2\pi}\int_0^{2\pi}{\rm
d}x~\delta u(x,t), \label{eq:deltar-incoherent-state} \\
&&\delta Z^\star(x,t)=\int_0^{2\pi}{\rm
d}y~G(x-y) \delta u(y,t).
\label{eq:deltaZj-incoherent-state}
\end{eqnarray}
Let us use the expansion $\delta u(x,t)=a(q)e^{iqx}e^{\lambda t}$,
with real $\lambda$, and with
the wave number $q$ being an integer (in view of having a periodic
spatial domain). Substituting in Eq.
(\ref{eq:u-expansion-incoherent-state}), and using $r(t)=1/(2\pi)\int_0^{2\pi}{\rm
d}x~u^\star(x,t)$, see Eq. (\ref{eq:r-exact}), we get
$r(t)=1/(2\pi)\int_0^{2\pi} {\rm d}x~ a^\star(q)e^{iqx}e^{\lambda
t}=a^\star(q=0)e^{\lambda t}$, so  that $r(t)$ being real implies that
$a(q)$ has to be real. On using $\delta u(x,t)=a(q)e^{iqx}e^{\lambda
t}$ in Eq. (\ref{eq:u-differential-equation-incoherent-state-linearized}), we get the
spectral equation determining the parameter $\lambda$:
\begin{equation}
\lambda=-\Delta+\frac{\delta_{q,0}}{2}+\frac{K\widetilde{G}(q)}{2}\equiv
\lambda(q),
\label{eq:lambda-incoherent-state}
\end{equation}
where  
\begin{equation}
\widetilde{G}(q)\equiv \int_0^{2\pi}{\rm d}y~G(x-y)
e^{iq(y-x)}=\int_0^{2\pi}{\rm d}z~G(z) e^{-iqz}=\frac{\sin(2\pi
q\sigma)}{2\pi q \sigma}
\label{eq:G-Fourier-transform}
\end{equation}
is the Fourier transform of the coupling function $G$; note that we have $\widetilde{G}(0)=1$. In arriving at Eq.
(\ref{eq:G-Fourier-transform}), we have
used the fact that $G$ is an even function of its argument,
$G(x)=G(-x)$, see Eq.
(\ref{eq:G-definition}). 

\begin{figure}[!ht]
\centering
\includegraphics[width=160mm]{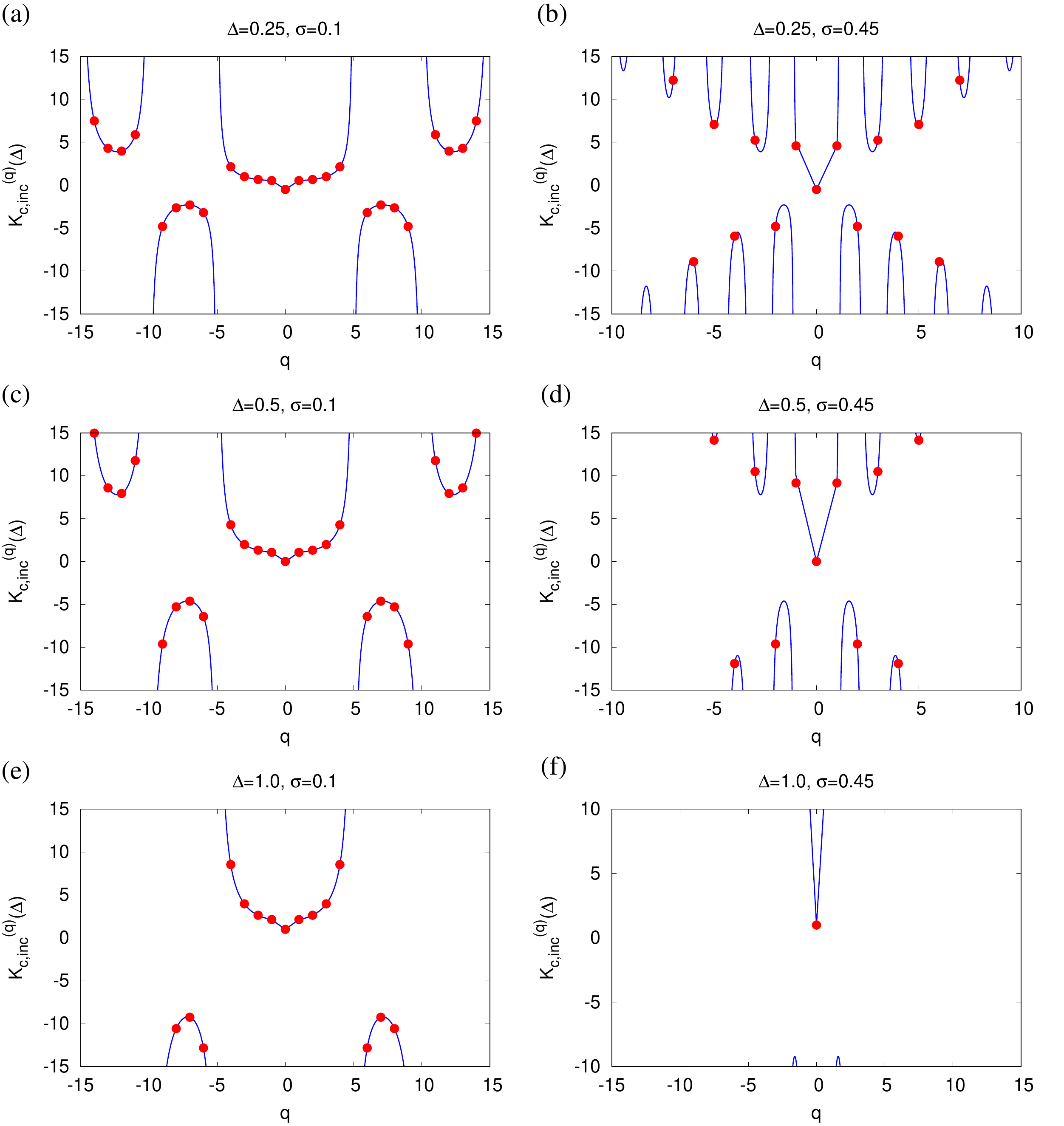}
\caption{Stability boundaries of the perturbation $\sim e^{iqx}$, see
Eq. (\ref{eq:stability-boundaries-incoherent-state}); at a fixed $q$, the perturbation grows in time for values of $K$ inside the
bounded region. For each panel, the values of $\Delta$ and $\sigma$ are
indicated in the figure. Here, the red dots are obtained by evaluating
Eq. (\ref{eq:stability-boundaries-incoherent-state}) for integer $q$,
and represent values relevant for our model, while the blue lines,
obtained by evaluating Eq.
(\ref{eq:stability-boundaries-incoherent-state}) for real $q$, serve as
a guide to the eye.}
\label{fig:stability-incoherent-state}
\end{figure}

Now, note that $\lambda(q)$ in Eq.
(\ref{eq:lambda-incoherent-state}) is real. Depending on whether
$\lambda(q)$ is larger or smaller than zero makes the perturbation $\sim
e^{iqx}$ grow or decay in time, respectively; the threshold between
the two behaviors is obtained by setting $\lambda(q)$ to zero in Eq.
(\ref{eq:lambda-incoherent-state}), thereby obtaining for a fixed
$\Delta$ the threshold
\begin{equation}
K_{c,{\rm
inc}}^{(q)}(\Delta)=\frac{2\Delta-\delta_{q,0}}{\widetilde{G}(q)}. 
\label{eq:stability-boundaries-incoherent-state}
\end{equation}
In particular, one has $K_{c, {\rm inc}}^{(0)}(\Delta)=2\Delta-1$. Equation (\ref{eq:lambda-incoherent-state})
may be rewritten as
\begin{equation}
\lambda=\frac{(K-K_{c, {\rm inc}}^{(q)}(\Delta))\widetilde{G}(q)}{2},
\label{eq:lambda-incoherent-state-delta}
\end{equation}
which implies that for values of $q$ such that $\widetilde{G}(q)>0$, the
perturbation $\sim e^{iqx}$ grows and is thus
sustained (respectively, decays, and is thus non-sustained) in time for $K > K_{c, {\rm inc}}^{(q)}(\Delta)$ (respectively,
$K < K_{c, {\rm inc}}^{(q)}(\Delta)$). On the other hand, for values of
$q$ such that $\widetilde{G}(q)<0$, the perturbation $\sim e^{iqx}$
grows and is thus
sustained (respectively, decays, and is thus non-sustained) in time for
$K < K_{c, {\rm inc}}^{(q)}(\Delta)$ (respectively,
$K > K_{c, {\rm inc}}^{(q)}(\Delta)$). For values of $q$ such that
$\widetilde{G}(q)=0$, Eq.
(\ref{eq:stability-boundaries-incoherent-state}) yields infinite
$K_{c,{\rm inc}}^{(q)}$; these
modes therefore do not exist for any finite $K$. The boundaries $K_{c, {\rm inc}}^{(q)}(\Delta)$
are shown in Fig. \ref{fig:stability-incoherent-state}.
From the figure, it is evident that the uniformly incoherent
state is stable with respect to perturbations $\sim e^{iqx}~\forall~q$
so long as $|K|$ is small and lies in the central region around zero that
has no overlap with any of the bounded regions. In fact for $K>0$, the incoherent
state destabilizes as a whole when it destabilizes with respect to
perturbations with wave number $q=0$. On increasing $K$, the uniformly incoherent state
becomes unstable with respect to a long-wavelength perturbation ($q=0$)
at $K=2\Delta-1$. On decreasing $K$, however, the instability is with
respect to a perturbation with a shorter wavelength ($q\ne0$).
%

\subsection{Synchronized twisted state}
\label{sec:synchronized-twisted-state}

\subsubsection{Existence}
\label{sec:synchronized-twisted-state-existence}

Let us look for solutions of Eq. (\ref{eq:u-differential-equation})
given by plane waves, or, the so-called (partially)
synchronized uniformly twisted states \cite{Omelchenko:2014}, which has the form
$u(x,t)=ae^{i(qx+\nu t)}$, with real $a,\nu$, and integer $q$. Here, the
wave number $q$ characterizes the ``twist" of the state, giving the rate of change of angle with $x$ at a fixed $t$, while $\nu$ measures the temporal rate of
rotation of the twisted state. The parameter $a$ with $0 < a < 1$
measures the level of coherence between the oscillator angles. We now obtain the conditions on the parameters $a,q,\nu$ for such a twisted state to be
a solution of Eq. (\ref{eq:u-differential-equation}). 
Substituting $u(x,t)=ae^{i(qx+\nu t)}$ in Eq. (\ref{eq:u-differential-equation}) gives
\begin{eqnarray}
&&i\nu +\Delta -\frac{1}{4\pi}\int_0^{2\pi}{\rm
d}x'~e^{iq(x'-x)}-\frac{K}{2}\int_0^{2\pi}{\rm
d}y~G(x-y)e^{iq(y-x)}
\nonumber \\
&&+\frac{a^2}{4\pi}\int_0^{2\pi}{\rm
d}x'~e^{-iq(x'-x)}+\frac{Ka^2}{2}\int_0^{2\pi}{\rm
d}y~G(x-y)e^{-iq(y-x)}=0.
\label{eq:u-differential-equation-synchronized-state}
\end{eqnarray}
Equating for real and imaginary
parts from both sides of Eq.
(\ref{eq:u-differential-equation-synchronized-state}), and using Eq. (\ref{eq:G-Fourier-transform}) and $\int_0^{2\pi}{\rm
d}x'~e^{iq(x'-x)}=2\pi \delta_{q,0}$, we get
\begin{eqnarray}
&&\Delta +\frac{a^2-1}{2}\delta_{q,0}+\frac{K(a^2-1)\widetilde{G}(q)}{2}=0,
\label{eq:u-differential-equation-synchronized-state-real} \\
&&\nu=0.
\label{eq:u-differential-equation-synchronized-state-imaginary}
\end{eqnarray}
The second equation implies that the synchronized uniformly
twisted state is actually a stationary solution of Eq.
(\ref{eq:u-differential-equation}). 
On the other hand, the first equation gives
\begin{equation}
a^2=1-\frac{2\Delta}{K\widetilde{G}(q)}\equiv
a^2(q);~~q\ne0,
\label{eq:twisted-state-a-condition}
\end{equation}
and
\begin{equation}
a^2(q=0)=1-\frac{2\Delta}{1+K}.
\label{eq:twisted-state-a-condition-1}
\end{equation}
Requiring $a(q=0)$ to be real implies that a zero-twist state is a stationary
solution of Eq. (\ref{eq:u-differential-equation}) provided that for fixed
$\Delta$, the parameter $K$ is larger than the critical value
$K_c^{(0)}\equiv 2\Delta-1$. On the other hand, Eq.
(\ref{eq:twisted-state-a-condition}) implies that a $q \ne 0$-twist
state is stationary for $K > K_c^{(q)}\equiv 2\Delta/\widetilde{G}(q)$
for $\widetilde{G}(q)>0$ and for $K < K_c^{(q)}$
for $\widetilde{G}(q)<0$. Comparing with Eq.
(\ref{eq:stability-boundaries-incoherent-state}), we see that
$K_c^{(q)}=K_{c,{\rm inc}}^{(q)}$, and thus, the instability boundaries
for the uniformly incoherent state shown in Fig.
\ref{fig:stability-incoherent-state} are also the existence boundaries
for the twisted states. In other words, the twisted state of wave number
$q$ emerges as the incoherent state destabilizes with respect to
perturbations of the same wave number. When exists, the synchronized uniformly twisted stationary state is represented as
$u^{\rm syn}_{\rm st}(x)=a(q)e^{iqx}$, with $a(q)$ given by Eqs.
(\ref{eq:twisted-state-a-condition}) and
(\ref{eq:twisted-state-a-condition-1}).

Corresponding to the zero-twist stationary state,
one has the stationary value $r^{\rm st}=a(q=0)$, while a $q\ne 0$-twist
state yields the stationary value $r^{\rm st}=0$,
where $a(q=0)$ is given by Eq. (\ref{eq:twisted-state-a-condition-1}). On the
other hand, Eq. (\ref{eq:Zj-exact}) yields the stationary value
\begin{equation}
Z^{\rm st}(x)=a(q)\int_0^{2\pi}{\rm
d}y~G(x-y)e^{-iqy}=a(q)e^{-iqx}\widetilde{G}(q),
\label{eq:Zj-synchronized-state}
\end{equation}
with $a(q)$ given by Eq. (\ref{eq:twisted-state-a-condition}).

\subsubsection{Stability}
\label{sec:synchronized-twisted-state-stability}

In this section, we study the linear stability of the twisted stationary
state, whose existence has been considered in Section
\ref{sec:synchronized-twisted-state-existence}. To this end, consider a twisted state with a given wave number $q_0$:
$u^{\rm syn}_{\rm st}(x)=a(q_0)e^{iq_0x}$. Using $r^{\rm
st}=a(q_0=0)\delta_{q_0,0}$, $Z^{\rm
st}(x)=a(q_0)e^{-iq_0x}\widetilde{G}(q_0)$, with $a(q_0)$ and $a(q_0=0)$
given respectively by Eqs. (\ref{eq:twisted-state-a-condition}) and
(\ref{eq:twisted-state-a-condition-1}), and writing $u(x,t)$ as 
\begin{equation}
u(x,t)=u^{\rm syn}_{\rm st}(x)+\delta u(x,t);~~|\delta u(x,t)|\ll 1,
\label{eq:u-expansion-synchronized-state}
\end{equation}
Eq. (\ref{eq:u-differential-equation}) yields to
leading order in $\delta u$ the equation
\begin{eqnarray}
&&\frac{\partial \delta u(x,t)}{\partial
t}+\Delta~\delta u+\frac{1}{2}\Big[2a(q_0=0)\delta_{q_0,0}u^{\rm syn}_{\rm
st}(x)\delta
u(x,t)\nonumber \\
&&+[u^{\rm syn}_{\rm st}(x)]^2\delta r(t)-\delta r^\star(t)\Big]\nonumber \\
&&+\frac{K}{2}\Big[2a(q_0)e^{-iq_0x}\widetilde{G}(q_0)u^{\rm syn}_{\rm
st}(x)\delta
u(x,t)\nonumber \\
&&+[u^{\rm syn}_{\rm st}(x)]^2\delta
Z(x,t)-\delta Z^\star(x,t)\Big]=0,
\label{eq:u-differential-equation-synchronized-state-linearized}
\end{eqnarray}
where we have
\begin{eqnarray}
&&\delta r^\star(t) \equiv \frac{1}{2\pi}\int_0^{2\pi}{\rm
d}x~\delta u(x,t), \label{eq:deltar-synchronized-state} \\
&&\delta Z^\star(x,t) \equiv \int_0^{2\pi}{\rm
d}y~G(x-y) \delta u(y,t).
\label{eq:deltaZj-synchronized-state}
\end{eqnarray}
Substituting $u^{\rm syn}_{\rm st}(x)=a(q_0)e^{iq_0x}$, and introducing $\delta U(x,t)\equiv\delta
u(x,t)e^{-iq_0x}$, Eq. (\ref{eq:u-differential-equation-synchronized-state-linearized})
yields
\begin{eqnarray}
&&\frac{\partial \delta U(x,t)}{\partial
t}+\left[\Delta+a^2(q_0)\Big(\delta_{q_0,0}+K\widetilde{G}(q_0)\Big)\right]\delta
U(x,t)\nonumber \\
&&+\frac{1}{2}\left[\frac{a^2(q_0)}{2\pi}\int_0^{2\pi}{\rm
d}y~\delta U^\star(y,t)e^{iq_0(x-y)}-\frac{1}{2\pi}\int_0^{2\pi}{\rm
d}y~\delta U(y,t)e^{iq_0(y-x)}\right]\nonumber \\
&&+\frac{K}{2}\Big[a^2(q_0)\int_0^{2\pi}{\rm
d}y~G(x-y) \delta U^\star(y,t)e^{iq_0(x-y)}\nonumber \\
&&-\int_0^{2\pi}{\rm
d}y~G(x-y) \delta U(y,t)e^{iq_0(y-x)}\Big]=0. 
\label{eq:u-differential-equation-linear-synchronized-state-linearized-1}
\end{eqnarray}

Let us introduce the column matrix
\begin{equation}
V(x,t)\equiv
\left( \begin{array}{c}
{\rm Re}~\delta U(x,t)  \\
{\rm Im}~\delta U(x,t)  \end{array} \right),
\label{eq:Vmatrix-definition}
\end{equation}
in terms of which Eq.
(\ref{eq:u-differential-equation-linear-synchronized-state-linearized-1})
may be rewritten as
\begin{eqnarray}
&&\frac{\partial V(x,t)}{\partial
t}+MV(x,t)\nonumber \\
&&+\frac{1}{2}\Big[\frac{a^2(q_0)}{2\pi}\int_0^{2\pi}{\rm
d}y~\left( \begin{array}{cc}
\cos[q_0(x-y)]&\sin[q_0(x-y)]  \\
\sin[q_0(x-y)]&-\cos[q_0(x-y)]  \end{array} \right) V(y,t)\nonumber \\
&&-\frac{1}{2\pi}\int_0^{2\pi}{\rm
d}y~\left( \begin{array}{cc}
\cos[q_0(x-y)]&\sin[q_0(x-y)]  \\
-\sin[q_0(x-y)]&\cos[q_0(x-y)]  \end{array} \right)V(y,t)\Big]\nonumber \\
&&+\frac{K}{2}\Big[a^2(q_0)\int_0^{2\pi}{\rm
d}y~G(x-y) \left( \begin{array}{cc}
\cos[q_0(x-y)]&\sin[q_0(x-y)]  \\
\sin[q_0(x-y)]&-\cos[q_0(x-y)]  \end{array} \right) V(y,t)\nonumber \\
&&-\int_0^{2\pi}{\rm
d}y~G(x-y) \left( \begin{array}{cc}
\cos[q_0(x-y)]&\sin[q_0(x-y)]  \\
-\sin[q_0(x-y)]&\cos[q_0(x-y)]  \end{array} \right)V(y,t)\Big]=0, 
\label{eq:u-differential-equation-linear-synchronized-state-linearized-2}
\end{eqnarray}
with
\begin{eqnarray}
M\equiv \left( \begin{array}{cc}
\Delta+a^2(q_0)\left(\delta_{q_0,0}+K\widetilde{G}(q_0)\right) & 0\\
0& \Delta+a^2(q_0)\left(\delta_{q_0,0}+K\widetilde{G}(q_0)\right)
\end{array} \right). \nonumber \\ 
\label{eq:Mmatrix-definition}
\end{eqnarray}

Next, we seek solutions to Eq.
(\ref{eq:u-differential-equation-linear-synchronized-state-linearized-2})
of the form
\begin{equation}
V(x,t)=\left(V_0 e^{iqx}+V^\star_0 e^{-iqx}\right)e^{\lambda
t}.
\label{eq:Vequation-solution}
\end{equation}
Substituting in Eq.
(\ref{eq:u-differential-equation-linear-synchronized-state-linearized-2}),
and using the identities
\begin{eqnarray}
&&\int_0^{2\pi}{\rm
d}y~\cos[q_0(x-y)]e^{iqy}=\pi\left(\delta_{q,-q_0}+\delta_{q,q_0}\right)e^{iqx},
\\
&&\int_0^{2\pi}{\rm
d}y~\sin[q_0(x-y)]e^{iqy}=i\pi\left(\delta_{q,-q_0}-\delta_{q,q_0}\right)e^{iqx},
\\
&&\int_0^{2\pi}{\rm
d}y~G(x-y)\cos[q_0(x-y)]e^{iqy}=\frac{1}{2}\left(\widetilde{G}(q+q_0)+\widetilde{G}(q-q_0)\right)e^{iqx},
\\
&&\int_0^{2\pi}{\rm
d}y~G(x-y)\sin[q_0(x-y)]e^{iqy}=\frac{i}{2}\left(\widetilde{G}(q+q_0)-\widetilde{G}(q-q_0)\right)e^{iqx},
\label{eq:G-identities}
\end{eqnarray}
we get
\begin{eqnarray}
&&\lambda \mathbb{I} V(x,t)+MV(x,t)\nonumber \\
&&+\frac{1}{4}\Big[a^2(q_0)\left( \begin{array}{cc}
g_+(q,q_0)&ig_-(q,q_0)  \\
ig_-(q,q_0)&-g_+(q,q_0)  \end{array} \right) \nonumber \\
&&-\left( \begin{array}{cc}
g_+(q,q_0)&ig_-(q,q_0)  \\
-ig_-(q,q_0)&g_+(q,q_0)  \end{array} \right)\Big]V(x,t)\nonumber \\
&&+\frac{K}{4}\Big[a^2(q_0) \left( \begin{array}{cc}
h_+(q,q_0)&ih_-(q,q_0)  \\
ih_-(q,q_0)&-h_+(q,q_0)  \end{array} \right) \nonumber \\
&&- \left( \begin{array}{cc}
h_+(q,q_0)&ih_-(q,q_0)  \\
-ih_-(q,q_0)&h_+(q,q_0)  \end{array} \right)\Big]V(x,t)=0, 
\label{eq:u-differential-equation-linear-synchronized-state-linearized-3}
\end{eqnarray}
with $\mathbb{I}$ being the $2\times 2$ identity matrix, and
\begin{eqnarray}
&&g_+(q,q_0) \equiv \delta_{q,-q_0}+\delta_{q,q_0}, \\
&&g_-(q,q_0) \equiv \delta_{q,-q_0}-\delta_{q,q_0}, \\
&&h_+(q,q_0) \equiv \widetilde{G}(q+q_0)+\widetilde{G}(q-q_0), \\
&&h_-(q,q_0) \equiv \widetilde{G}(q+q_0)-\widetilde{G}(q-q_0).
\label{eq:g+-h+-definitions}
\end{eqnarray}
On requiring the existence of the solution
(\ref{eq:Vequation-solution}), we obtain from Eq.
(\ref{eq:u-differential-equation-linear-synchronized-state-linearized-3}) the secular equation determining $\lambda$:
\begin{equation}
{\rm Det}(\lambda \mathbb{I}-B(q))=0,
\label{eq:eigenvalue-equation-synchronized-state}
\end{equation}
where we have
\begin{eqnarray}
&&B(q) \equiv -M-\frac{1}{4}a^2(q_0) \left( \begin{array}{cc}
g_+(q,q_0)&ig_-(q,q_0)  \\
ig_-(q,q_0)& -g_+(q,q_0)  \end{array} \right) \nonumber \\
&&+\frac{1}{4} \left( \begin{array}{cc}
g_+(q,q_0)&ig_-(q,q_0)  \\
-ig_-(q,q_0)& g_+(q,q_0)  \end{array} \right) \nonumber \\
&&-\frac{K}{4}a^2(q_0) \left( \begin{array}{cc}
h_+(q,q_0)&ih_-(q,q_0)  \\
ih_-(q,q_0)& -h_+(q,q_0)  \end{array} \right) \nonumber \\
&&+\frac{K}{4} \left( \begin{array}{cc}
h_+(q,q_0)&ih_-(q,q_0)  \\
-ih_-(q,q_0)& h_+(q,q_0)  \end{array} \right).
\label{eq:Bmatrix-definition}
\end{eqnarray}
The solutions of Eq. (\ref{eq:eigenvalue-equation-synchronized-state}) may be written
as
\begin{equation}
\lambda_\pm (q)=\frac{1}{2}\left({\rm Tr}(B(q))\pm\sqrt{[{\rm Tr}(B(q))]^2-4{\rm
Det}(B(q))}\right).
\label{eq:lambaplusminus-synchronized-state}
\end{equation}

From Eq. (\ref{eq:Bmatrix-definition}), it follows that
\begin{eqnarray}
&&\hspace{-3cm}B(q) =\nonumber \\
&&\hspace{-3cm}\left( \begin{array}{cc}
-\Delta-a^2(q_0)\left(\delta_{q_0,0}+K\widetilde{G}(q_0)\right)& -\frac{(a^2(q_0)-1)}{4}ig_-(q,q_0) \\
-\frac{(a^2(q_0)-1)}{4}g_+(q,q_0)-\frac{K(a^2(q_0)-1)}{4}h_+(q,q_0)&-\frac{K(a^2(q_0)-1)}{4}ih_-(q,q_0)  \\
& \\
-\frac{(a^2(q_0)+1)}{4}ig_-(q,q_0) &
-\Delta-a^2(q_0)\left(\delta_{q_0,0}+K\widetilde{G}(q_0)\right)\\
-\frac{K(a^2(q_0)+1)}{4}ih_-(q,q_0) &
+\frac{(a^2(q_0)+1)}{4}g_+(q,q_0)+\frac{K(a^2(q_0)+1)}{4}h_+(q,q_0)\\
\end{array} \right), \nonumber \\
\label{eq:Bmatrix-explicit}
\end{eqnarray} 
and hence, we get
\begin{eqnarray}
{\rm
Tr}(B(q))=-2\left(\Delta+a^2(q_0)\left(\delta_{q_0,0}+K\widetilde{G}(q_0)\right)\right)+\frac{g_+(q,q_0)}{2}+\frac{Kh_+(q,q_0)}{2}.
\nonumber \\
\label{eq:traceB}
\end{eqnarray}

Choosing $K > K_{c,{\rm inc}}^{(q_0)}=2\Delta/\widetilde{G}(q_0)$ for
$\widetilde{G}(q_0)>0$ or $K < K_{c,{\rm inc}}^{(q_0)}=2\Delta/\widetilde{G}(q_0)$ for $\widetilde{G}(q_0)<0$, so that the twisted state $\sim
e^{iq_0x}$ under consideration exists in the first place, we may evaluate $\lambda_\pm(q)$
by using Eq. (\ref{eq:lambaplusminus-synchronized-state}); having
a positive (respectively, negative) $\lambda_\pm(q)$ implies that the
perturbation $\sim e^{iqx}$ to the twisted state grows (respectively,
decays) and is thus sustained (respectively, non-sustained) in time.
Note that we have $\lambda_+(q) > \lambda_-(q)$.

\subsubsection{Stability of the zero-twist state}
\label{sec:synchronized-twisted-state-stability-q0-0}

For the particular case $q_0=0,q\ne 0$, we need to choose $K > K_{c,{\rm inc}}^{(0)}=2\Delta-1$. In this case, Eq.
(\ref{eq:Bmatrix-explicit})
reduces to
\begin{eqnarray}
\hspace{-2cm}B(q\ne 0) = \left( \begin{array}{cc}
-\Delta-a^2(0)\left(1+K\right)&0 \\
-\frac{K(a^2(0)-1)}{2}\widetilde{G}(q)&  \\
& \\
0 &
-\Delta-a^2(0)\left(1+K\right)\\
 &
+\frac{K(a^2(0)+1)}{2}\widetilde{G}(q)\\
\end{array} \right), 
\label{eq:Bmatrix-explicit-q0-0-qnot0}
\end{eqnarray} 
so that
\begin{eqnarray}
&&{\rm
Tr}(B(q\ne
0))=2(\Delta-1-K)+K\widetilde{G}(q),\nonumber \\
\label{eq:traceB-q0-0-qnot0} \\
&&{\rm Det}(B(q\ne
0))=\left(1+K-\Delta-\frac{K\widetilde{G}(q)}{2}\right)^2-\frac{K^2a^4(0)[\widetilde{G}(q)]^2}{4},
\nonumber
\end{eqnarray}
and hence, 
\begin{equation}
\lambda_\pm(q \ne 0)=\frac{2(\Delta-1-K)+K\widetilde{G}(q) (1\mp a^2(0))}{2}.
\label{eq:q0-0-qnot0-lambdap-lambdam} 
\end{equation}
Using Eq. (\ref{eq:twisted-state-a-condition-1}) and the fact that $K >
2\Delta-1$, and that $\widetilde{G}(q\ne 0)<1$, we may write
$\lambda_+(q \ne 0)< \Lambda \equiv 
-\Delta+K\frac{\Delta}{1+K}$; simplifying, we get
$\Lambda=-\frac{\Delta}{1+K}$. Using again $K >
2\Delta-1$, and noting that $\Delta$ is a positive quantity, it follows
that $1+K$ is also positive, and hence, $\Lambda$ is negative, implying
that $\lambda_+(q\ne 0)<0$.

On the other hand, for $q=0$, we have
\begin{eqnarray}
\hspace{-2cm}B(q) = \left( \begin{array}{cc}
2\Delta-K-1& 0 \\
& \\
0 &
0\\
\end{array} \right), 
\label{eq:Bmatrix-explicit-q0-0-q0}
\end{eqnarray} 
and hence, we get
\begin{equation}
{\rm
Tr}(B(q=0))=2\Delta-1-K < 0,~{\rm
Det}(B(q=0))=0,
\label{eq:Bmatrix-q0-0-q0-trace-det} 
\end{equation}
where we have used Eq. (\ref{eq:twisted-state-a-condition-1}) and the fact
that $K >
K_{c,{\rm inc}}^{(0)}=2\Delta-1$; we thus have
\begin{equation}
\lambda_+(0)=0,~\lambda_-(0)=2\Delta-1-K<0.
\label{eq:q0-0-q0-lambdap-lambdam} 
\end{equation}
Using the facts $\lambda_+(q\ne
0)<\lambda_+(0)$ and $\lambda_+(q) > \lambda_-(q)$, we conclude that the stability of the zero-twist state
is determined by the behavior of $\lambda_+(0)$. Namely, at a fixed
$\Delta$, and for $K > 2\Delta-1$, we have $\lambda_+(0)=0$, so that
referring to Fig. \ref{fig:stability-incoherent-state}, we see that the zero-twist state stabilizes as soon as
the incoherent one destabilizes. At a fixed $\Delta$, the transition
between the two states takes place at $K=K_{c,{\rm
inc}}^{(0)}=2\Delta-1$.

\subsection{Phase transition in the Kuramoto order parameter}
\label{sec:incoherent-zero-twist-synchronized-phase-transition}
On the basis of our discussions in Sections
\ref{sec:uniformly-incoherent-state-stability} and
\ref{sec:synchronized-twisted-state}, we conclude that the
stationary state Kuramoto order parameter $r^{\rm st}$ undergoes a
continuous transition at a fixed $\Delta$, from a low-$K$ zero value, corresponding
to an incoherent state, to a high-$K$ non-zero value $r^{\rm
st}=\sqrt{1-\frac{2\Delta}{1+K}}$, corresponding to a
zero-twist synchronized state, at the critical threshold
\begin{equation}
K_c(T=0,\Delta)=2\Delta-1.
\label{eq:phase-boundary-DK}
\end{equation}
The line of transition $K_c(T=0,\Delta)$ is shown in
the phase diagram, Fig. \ref{fig:schematic-phase-diagram}. The line
intercepts the $\Delta$-axis at the point $\Delta=1/2$, which matches
with the prediction for this point made on the basis of the analysis of
the bare Kuramoto model, that is, in the absence of any non-local
interactions, see the discussions in Section \ref{sec:phase-transition}. Note that
we may write $r^{\rm st}=\sqrt{\frac{K-K_c(T=0,\Delta)}{1+K}}$, from which we obtain as $K\to
K_c^+(T=0,\Delta)$ the scaling $r^{\rm st}\sim
\left(K-K_c(T=0,\Delta)\right)^\delta$, where the critical exponent $\delta$ has
the value $\delta=1/2$.

\section{Analysis for the $(K,T)$-plane with $\Delta=0$}
\label{sec:KT-analysis}

In this section, we discuss the phase diagram of the model
(\ref{eq:eom-general}) in the $(K,T)$-plane, i.e., for $\Delta=0$.
We first consider the case of finite $M$, present an analysis of the
phase diagram as a function of $M$, and, in the end, consider the
limit $M \to \infty$ of the results. 

As discussed in Section \ref{sec:model}, the
dynamics (\ref{eq:eom-general}) for $\Delta=0$ relaxes at long times to an equilibrium stationary state with the BG
distribution for the angles, Eq.
(\ref{eq:Delta0-equilibrium-distribution}). The canonical partition function
for a $1d$ periodic chain of $N$ sites is thus given by
\begin{eqnarray}
Z_N&=&\int\left(\prod_{j=1}^N {\rm d}\theta_j\right) \exp[-\beta {\cal V}(\{\theta_j\})] \nonumber \\
&=&e^{-\beta N/2}\int\left(\prod_{j=1}^N {\rm d}\theta_j\right)
\exp\Big[
\frac{\beta}{2N}\Big\{\Big(\sum_{j=1}^N\cos\theta_j\Big)^2+\Big(\sum_{j=1}^N\sin\theta_j\Big)^2\Big\}\nonumber
\\
&&+\frac{\beta K}{4M}\sum_{j=1}^N\sum_{k=-M}^M\cos(\theta_{j+k}-\theta_j)\Big],
\label{eq:canonical-partition-function}
\end{eqnarray}
with $\beta\equiv 1/T$. 

Next, using the Hubbard-Stratonovich transformation,
\begin{equation}
\exp(ax^2)=\frac{1}{\sqrt{4\pi a}}\int_{-\infty}^\infty {\rm
d}z~\exp\left(-\frac{z^2}{4a}+ zx\right);~~a>0,
\label{eq:hubbard-stratonovich-transformation}
\end{equation}
in Eq. (\ref{eq:canonical-partition-function}), and introducing auxiliary fields $\widetilde{z}_1 \equiv \beta z_1$ and
$\widetilde{z}_2 \equiv \beta z_2$, we obtain 
\begin{eqnarray}
Z_N&=&e^{-\beta N/2}\frac{N}{2\pi
\beta}\int_{-\infty}^\infty {\rm
d}\widetilde{z}_1 \int_{-\infty}^\infty
{\rm d}\widetilde{z}_2 \int \left(\prod_{j=1}^N {\rm d}\theta_j\right)
\exp\Big[-\frac{N}{2\beta}(\widetilde{z}_1^2+\widetilde{z}_2^2)\nonumber \\
&&+\widetilde{z}_1
\sum_{j=1}^N\cos \theta_j +\widetilde{z}_2\sum_{j=1}^N\sin
\theta_j+\frac{\beta K}{4M}\sum_{j=1}^N\sum_{k=-M}^M\cos(\theta_{j+k}-\theta_j)\Big]
\nonumber \\
&=&e^{-\beta N/2}\frac{N\beta}{2\pi}\int_{-\infty}^\infty {\rm d}z_1\int_{-\infty}^\infty
{\rm d}z_2 \int \left(\prod_{j=1}^N {\rm d}\theta_j\right)
\exp\Big[-\frac{N\beta}{2}(z_1^2+z_2^2)\nonumber \\
&&+\beta z_1
\sum_{j=1}^N\cos \theta_j +\beta z_2\sum_{j=1}^N\sin
\theta_j+\frac{\beta K}{4M}\sum_{j=1}^N\sum_{k=-M}^M\cos(\theta_{j+k}-\theta_j)\Big].
\label{eq:canonical-partition-function-auxiliary-fields}
\end{eqnarray}
Writing $z_1=z\cos \phi, z_2=z\sin \phi$, with real
$z=(z_1^2+z_2^2)^{1/2}>0$ and
$\phi \in [0,2\pi)$ given by $\phi=\tan^{-1}(z_2/z_1)$, we get
\begin{eqnarray}
Z_N&=&\frac{N\beta}{2\pi} \int_0^{2\pi}{\rm d}\phi~
\int_0^\infty{\rm d}z~ z\int \left(\prod_{j=1}^N {\rm d}\theta_j\right)
\exp\Big[-\frac{N\beta}{2}(1+z^2)\nonumber \\
&&+\beta z \sum_{j=1}^N\cos (\theta_j-\phi)+\frac{\beta K}{4M}\sum_{j=1}^N\sum_{k=-M}^M\cos(\theta_{j+k}-\theta_j)\Big].
\label{eq:canonical-partition-function-auxiliary-fields-phi}
\end{eqnarray}
In view of the invariance of the potential (\ref{eq:V}) under
rotation by an equal amount of all the $\theta_j$'s, we get
\begin{eqnarray}
Z_N&=&\frac{N\beta}{2\pi} \int_0^{2\pi}{\rm d}\phi~
\int_0^\infty{\rm d}z~ z\int \left(\prod_{j=1}^N {\rm d}\theta_j\right)
\exp\Big[-\frac{N\beta}{2}(1+z^2)\nonumber \\
&&+\beta z \sum_{j=1}^N\cos \theta_j+\frac{\beta
K}{4M}\sum_{j=1}^N\sum_{k=-M}^M\cos(\theta_{j+k}-\theta_j)\Big].
\label{eq:canonical-partition-function-2-5}
\end{eqnarray}

Below, we consider separately the cases $K=0$ and $K \ne 0$.

\subsection{$K=0$}
\label{sec:KT-analysis-K0}

For $K=0$, Eq. (\ref{eq:canonical-partition-function-2-5}) yields
\begin{eqnarray}
Z_N&=&N\beta\int_0^\infty 
{\rm d}z~z\exp\left[-\frac{N\beta}{2}(1+z^2)\right]\int
\left(\prod_{j=1}^N {\rm d}\theta_j\right)
~\exp\left[\beta z \sum_{j=1}^N\cos
\theta_j\right] \nonumber \\
&=&N\beta\int_0^\infty 
{\rm d}z~z\exp\left[-N\left\{\frac{\beta}{2}(1+z^2)-\ln \left(\int_0^{2\pi} {\rm d}\theta
~\exp(\beta z \cos \theta)\right)\right\}\right]. \nonumber \\
\label{eq:ZN-K0-explicit}
\end{eqnarray}
In the thermodynamic limit, $Z_N$ may be approximated by
invoking the saddle-point method to perform
the integration in $z$ on the right hand side; one gets
\begin{equation}
Z_N=N\beta~z_s\exp\left[-N\left\{\frac{\beta}{2}(1+z_s^2)-\ln \left(\int_0^{2\pi} {\rm d}\theta
~\exp(\beta z_s \cos \theta)\right)\right\}\right], 
\label{eq:ZN-K0-explicit-1}
\end{equation}
where the saddle-point value $z_s$ solves the equation
\begin{eqnarray}
z_s&=&\frac{\int_0^{2\pi} {\rm d}\theta
~\cos \theta\exp(\beta z_s \cos \theta)}{\int_0^{2\pi} {\rm d}\theta
~\exp(\beta z_s \cos \theta)} \nonumber \\
&=&\frac{I_1(\beta z_s)}{I_0(\beta z_s)},
\label{eq:saddle-point-K0}
\end{eqnarray}
where $I_n(x)=(1/(2\pi))\int_0^{2\pi}{\rm d}\theta~\exp(x\cos
\theta)\cos(n\theta)$ is the modified Bessel function of first kind and
order $n$. It may be shown that $z_s$ is nothing but the synchronization order parameter
$r^{\rm st}$, see Section \ref{sec:KT-analysis-Kneq0} below. Equation (\ref{eq:saddle-point-K0}) has a
trivial solution $r^{\rm st}=0$ valid at all temperatures, while a non-zero
solution exists for $\beta \ge \beta_c=2$ \cite{Campa:2014}. In fact,
the system shows a continuous transition, from a synchronized/magnetized
phase ($r^{\rm st}\ne 0$) at low temperatures to an incoherent/unmagnetized
phase ($r^{\rm st}=0$) at high temperatures at the critical temperature
$T_c=1/2$. The latter point coincides with the BMF phase transition point
$T_c(\Delta=0,K=0)$ indicated in the phase diagram in Fig.
\ref{fig:schematic-phase-diagram}; this is a consequence of the fact that the
BMF model and our model with $\Delta=K=0$ have the same distribution of
the angles in equilibrium given by Eqs.
(\ref{eq:Delta0-equilibrium-distribution}) and (\ref{eq:V}) with $K=0$ \cite{Chavanis:2014}.

\subsection{$K \ne 0$}
\label{sec:KT-analysis-Kneq0}

For $K \ne 0$, Eq. (\ref{eq:canonical-partition-function-2-5}) gives
\begin{eqnarray}
Z_N&=&N\beta\int_0^\infty 
{\rm d}z~z\exp\left[-\frac{N\beta}{2}(1+z^2)\right]{\cal
Z}_N;\label{eq:canonical-partition-function-3}\\
{\cal Z}_N&\equiv&\int \Big(\prod_{j=1}^N {\rm d}\theta_j\Big)
\exp\Big[\beta z \sum_{j=1}^N\cos
\theta_j+\frac{\beta K}{4M}\sum_{j=1}^N\sum_{k=-M}^M\cos(\theta_{j+k}-\theta_j)\Big].
\label{eq:canonical-partition-function-auxiliary-fields-nophi}
\end{eqnarray}
Here, we may identify the factor ${\cal Z}_N$ with the canonical partition function
of a $1d$ periodic chain of $N$ phase-only oscillators (equivalently, classical $XY$-spins), where each
oscillator interacts with strength $K/(4M)$ with $M$
neighboring oscillators to the left and to
the right, and also with an external field of strength $z$ along the $x$
direction. Note from Eq.
(\ref{eq:canonical-partition-function-auxiliary-fields-nophi}) that
under $z \to -z$, one has
\begin{eqnarray}
&&{\cal Z}_N\to{\cal Z}^\prime_N \nonumber \\
&&\equiv \int \left(\prod_{j=1}^N {\rm d}\theta_j\right)
\exp\left[-\beta z \sum_{j=1}^N\cos
\theta_j+\frac{\beta
K}{4M}\sum_{j=1}^N\sum_{k=-M}^M\cos(\theta_{j+k}-\theta_j)\right]
\nonumber \\
&&=\int \left(\prod_{j=1}^N {\rm d}\theta_j\right)
\exp\left[\beta z \sum_{j=1}^N\cos (\theta_j+\pi)+\frac{\beta
K}{4M}\sum_{j=1}^N\sum_{k=-M}^M\cos(\theta_{j+k}-\theta_j)\right],
\nonumber \\
\label{eq:canonical-partition-function-auxiliary-fields-nophi-ztominusz}
\end{eqnarray}
so that on using the invariance of the potential (\ref{eq:V}) under
rotation of all the $\theta_j$'s by an amount equal to $\pi$, we
obtain that ${\cal Z}^\prime_N={\cal Z}_N$. We thus conclude that the factor ${\cal Z}_N$ is an even function of $z$.

Invoking the above mentioned analogy with the $1d$ periodic chain of
oscillators, we now proceed to compute the factor ${\cal Z}_N$ for large $N$. Our approach is based on a combination of a matrix formulation that was
developed to study a general spin model in $1d$ with an $n$-neighbor
interaction, with $n$ arbitrary and finite \cite{Montroll:1942,Dobson:1969}, and a transfer operator
method that generalizes the well-known transfer matrix approach for Ising
spins \cite{Huang:1987} to the case of continuous spins \cite{Dauxois:2006}.

The starting point is to consider a $1d$ periodic chain of total number
of sites equal to ${\cal N}M$, and then to divide it into ${\cal N}$
blocks of $M$ sites. Let us relabel the
oscillators (equivalently, the sites accommodating them), such that
$\theta_j^{(\alpha)}$ refers to the angle of the $j$-th oscillator within the
$\alpha$-th block, with $\alpha=1,2,\ldots,{\cal N}$, and $j=1,2,\ldots,M$. By
virtue of such a construction, an oscillator in the $\alpha$-th block interacts 
with the oscillators in the same block and with those in the
$(\alpha-1)$-th and $(\alpha+1)$-th blocks. As a result, the total energy of any configuration of
the oscillator angles may be expressed as a sum of (i) energies due to
interaction of the oscillators with the external field of strength $z$, (ii) interaction energies of
oscillators within the same block, and (iii) interaction energies of oscillators
from adjacent blocks. Next, let us denote the configuration
of the $\alpha$-th block by $C_\alpha \equiv
\{\theta_j^{(\alpha)}\}_{1\le j \le M}$. With this notation, we may express the energy of the
system in configuration $C \equiv (C_1,C_2,\ldots,C_{\cal N})$ as 
\begin{equation}
H(C)=X_{C_1}+Y_{C_1,C_2}+X_{C_2}+Y_{C_2,C_3}+\ldots+Y_{C_{\cal N},C_1},
\label{eq:canonical-partition-function-HC}
\end{equation}
where $X_{C_\alpha}$ denotes the energy contribution due to interaction of types (i) and (ii), and
$Y_{C_\alpha,C_{\alpha+1}}$ denotes the energy due to interaction
of type (iii) contributed by the oscillators in the $\alpha$-th and
$(\alpha+1)$-th blocks:
\begin{eqnarray}
&&X_{C_\alpha}\equiv -z\sum_{j=1}^M \cos
\theta^{(\alpha)}_{j}-\frac{K}{2M}\sum_{j=1}^M\sum_{k=1}^{M-j}\cos(\theta^{(\alpha)}_{j+k}-\theta^{(\alpha)}_{j}),
\nonumber
\\
\label{eq:canonical-partition-function-XC-YC} \\
&&Y_{C_\alpha,C_{\alpha+1}}\equiv-\frac{K}{2M}\sum_{j=1}^M
\sum_{k=1}^j\cos(\theta^{(\alpha+1)}_{j-k+1}-\theta^{(\alpha)}_{M+1-k}).
\nonumber
\end{eqnarray}

The transfer operator method \cite{Dauxois:2006} introduces an operator
${\cal T}(C,C')$ as
\begin{equation}
{\cal T}(C,C')\equiv
\exp\left(-\beta\Big[\frac{1}{2}X_{C}+Y_{C,C'}+\frac{1}{2}X_{C'}\Big]\right).
\label{eq:TCC}
\end{equation}

Let $\{\lambda_q\}$ be the set of eigenvalues \footnote{Note that the
operator ${\cal
T}(C,C')$ is not symmetric in $(C,C')$, so that one has to
distinguish between its left and right eigenvalues and eigenvectors. Here, $\lambda_q$'s refer to the
set of the right eigenvalues of ${\cal T}(C,C')$.} of the transfer operator ${\cal
T}(C,C')$. In other words, denoting the
eigenfunctions of ${\cal
T}(C,C')$ as $f_q(C)$, we have
\begin{equation}
\int {\rm d}C'~{\cal T}(C,C')f_q(C')=\lambda_q f_q(C).
\label{eq:lambdaq-definition}
\end{equation}
In terms of $\{\lambda_q\}$, we obtain the canonical
partition for a $1d$ ring of ${\cal N}M$ sites as
\begin{eqnarray}
{\cal Z}_{{\cal N}M}&=&\int{\rm d}C_1{\rm
d}C_2\ldots{\rm d}C_{\cal N}~{\cal T}(C_1,C_2){\cal T}(C_2,C_3){\cal
T}(C_3,C_4)\ldots \nonumber \\
&&\times {\cal T}(C_{{\cal N}-2},C_{{\cal N}-1}){\cal T}(C_{{\cal
N}-1},C_{\cal N}){\cal T}(C_{\cal N},C_1) \nonumber \\
&=&\sum_q \left[\lambda_q\Big(\beta z,\frac{\beta K}{M}\Big)\right]^{\cal N},
\label{eq:canonical-partition-function-lambda}
\end{eqnarray}
where we have $\int {\rm d}C_\alpha\equiv \int (\prod_{j=1}^M) {\rm
d}\theta^{(\alpha)}_{j}$. For large ${\cal N}$, the sum in Eq. (\ref{eq:canonical-partition-function-lambda})
is dominated by the largest eigenvalue $\lambda_{\rm
max}=\lambda_{\rm max}\left(\beta z,\frac{\beta K}{M}\right)$,
yielding \footnote{For any infinitesimal discretization of the
$\theta_j$'s, the operator ${\cal T}(C,C')$ becomes a finite-dimensional
real square matrix
with positive entries, so that the
application of the Perron-Frobenius theorem \cite{Perron-Frobenius} implies the existence of the
largest eigenvalue that is real and non-degenerate.} 
\begin{equation}
{\cal Z}_{{\cal N}M}=\lambda_{\rm max}^{\cal N}.
\label{eq:canonical-partition-function-lambdamax}
\end{equation}
For our system of interest, Eq. (\ref{eq:eom-general}), we have ${\cal N}M=N$, giving  
\begin{equation}
{\cal Z}_{N}=\lambda_{\rm max}^{N/M}.
\label{eq:canonical-partition-function-lambdamax-again}
\end{equation}

Equation (\ref{eq:canonical-partition-function-lambdamax-again}) when combined with the fact shown earlier that ${\cal
Z}_N$ is an even function of
$z$ implies that $\lambda_{\rm max}$ is an even function of $z$.
Substituting Eq. (\ref{eq:canonical-partition-function-lambdamax-again})
in Eq.  
(\ref{eq:canonical-partition-function-3}), we
obtain in the thermodynamic limit the result
\begin{equation}
Z_N=N\beta\int_0^\infty 
{\rm d}z~z\exp\left[-N\Big\{\frac{\beta}{2}(1+z^2)-\frac{1}{M}\ln \lambda_{\rm
max}\Big(\beta
z,\frac{\beta K}{M}\Big)\Big\}\right].
\label{eq:canonical-partition-function-full}
\end{equation}
In the same limit, one may further approximate $Z_N$ by
invoking the saddle-point method to perform
the integration in $z$; one gets
\begin{equation}
Z_N=N\beta z_s\exp\left[-N\Big\{\frac{\beta}{2}(1+z_s^2)-\frac{1}{M}\ln \lambda_{\rm
max}\Big(\beta
z_s,\frac{\beta K}{M}\Big)\Big\}\right],
\label{eq:canonical-partition-function-saddle-point}
\end{equation}
where $z_s$ solves the saddle-point equation
\begin{equation}
z_s\equiv\sup_z \widetilde{\phi}(\beta,z), 
\label{eq:saddle-point-definition-1}
\end{equation}
with $\widetilde{\phi}(\beta,z)$ being the free-energy function:
\begin{equation}
-\widetilde{\phi}(\beta,z)\equiv-\frac{\beta}{2}(1+z^2)+\frac{1}{M}\ln
\lambda_{\rm max}\left(\beta
z,\frac{\beta K}{M}\right). \label{eq:saddle-point-definition-2}
\end{equation}
The saddle-point equation may thus be written as 
\begin{equation}
z_s=\frac{1}{M}\frac{\partial \ln \lambda_{\rm max}\left(\beta z,\frac{\beta
K}{M}\right)}{\partial(\beta z)}\Big|_{z=z_s}.
\label{eq:saddle-point-equation}
\end{equation}
From Eq. (\ref{eq:canonical-partition-function-saddle-point}), one
obtains the dimensionless free energy per
oscillator, $\phi(\beta)\equiv -\lim_{N \to \infty} (\ln Z_N)/N$, as
\begin{equation}
-\phi(\beta)=\sup_z \left[-\widetilde{\phi}(\beta,z)\right],
\label{eq:phi-beta}
\end{equation}
where we have suppressed the dependence of $\phi(\beta)$ on $K$. We thus
have 
\begin{equation}
-\phi(\beta)\equiv-\frac{\beta}{2}(1+z_s^2)+\frac{1}{M}\ln \lambda_{\rm
max}\left(\beta
z_s,\frac{\beta K}{M}\right).
\label{eq:phi-beta-final}
\end{equation}
Note that the free energy at a given temperature has a definite value
given by Eq. (\ref{eq:phi-beta-final}), and is obtained by substituting the saddle-point solution $z_s$ into the
expression for the free-energy function $\widetilde{\phi}(\beta,z)$. 

As it turns out, the quantity $z_s$ in Eq.
(\ref{eq:saddle-point-equation}) is nothing but the stationary Kuramoto order parameter
$r^{\rm st}$. To demonstrate that this is the case, consider the dynamics (\ref{eq:eom-general})
with $\Delta=0$ and in presence of an
additional potential $V_{\rm ext}(\{\theta_j\})\equiv -h\sum_{j=1}^N
\cos \theta_j$ due to an external field of strength $h$ along the
$x$-direction, so that the partition function
(\ref{eq:canonical-partition-function}) is modified to
$Z_N^{(h)}\equiv\int\left(\prod_{j=1}^N {\rm d}\theta_j\right) \exp[-\beta
\{{\cal V}(\{\theta_j\})+V_{\rm ext}(\{\theta_j\})\}]$. In this case,
one obtains in the same way as one arrives at Eqs.
(\ref{eq:saddle-point-equation}) and
(\ref{eq:phi-beta-final}) the following
analogous equations
\begin{eqnarray}
&&z_s=\frac{1}{M}\frac{\partial \ln \lambda_{\rm max}\left(\beta (z+h),\frac{\beta
K}{M}\right)}{\partial \left(\beta(z+h)\right)}\Big|_{z=z_s},
\label{eq:saddle-point-equation-h} \\
&&-\phi(\beta,h)=-\frac{\beta}{2}(1+z_s^2)+\frac{1}{M}\ln
\lambda_{\rm max}\left(\beta (z_s+h),\frac{\beta K}{M}\right),
\label{eq:phi-beta-final-h}
\end{eqnarray}
where note that $z_s$ in Eq. (\ref{eq:phi-beta-final-h}) is a function
of $\beta h$ and $\beta K/(4M)$ by virtue of Eq.
(\ref{eq:saddle-point-equation-h}). On the other hand, the stationary
Kuramoto order parameter in presence of the field $h$ has values
$r_x^{\rm st}(h)\ne 0$ and $r_y^{\rm st}(h)=0$, so that one obtains for
$r^{\rm st}(h)=r_x^{\rm st}(h)$ in the thermodynamic limit  
\begin{eqnarray}
r^{\rm st}(h)&=&\lim_{N\to\infty}\frac{1}{NZ_N^{(h)}}\int\left(\prod_{j=1}^N {\rm
d}\theta_j\right)\left(\sum_{l=1}^N \cos \theta_l \right) \nonumber \\
&&\times \exp[-\beta
\{{\cal V}(\{\theta_j\})+V_{\rm ext}(\{\theta_j\})\}] \nonumber \\
&=&\lim_{N\to \infty}\frac{1}{N}\frac{\partial \ln Z_N^{(h)}}{\partial (\beta h)}\nonumber
\\
&=&-\lim_{N\to \infty}\frac{\partial \phi(\beta,h)}{\partial (\beta h)} \nonumber \\
&=&-\beta z_s \frac{\partial z_s}{\partial (\beta h)} +
\frac{1}{M}\frac{\partial \ln \lambda_{\rm max}\left(\beta (z+h),\frac{\beta
K}{M}\right)}{\partial \left(\beta (z+h)\right)}\Big|_{z=z_s} \left[\beta \frac{\partial
z_s}{\partial (\beta h)} + 1\right] \nonumber \\
&=&\frac{1}{M}\frac{\partial \ln \lambda_{\rm max}\left(\beta (z+h),\frac{\beta
K}{M}\right)}{\partial \left(\beta(z+h)\right)}\Big|_{z=z_s},
\label{eq:rst-h}
\end{eqnarray}
where in obtaining the third equality, we have used the result that
$\phi(\beta,h)=-\lim_{N \to \infty}(\ln Z_N^{(h)}))/N$, while in
obtaining the last two equalities, we have used Eqs. (\ref{eq:saddle-point-equation-h}) and (\ref{eq:phi-beta-final-h}). 
Comparing Eqs. (\ref{eq:saddle-point-equation-h}) and (\ref{eq:rst-h}),
we conclude that $r^{\rm st}(h)=z_s$; It is evident from the
derivation of this result that it holds for all values of $h$, including
$h=0$. We have thus established the assertion made above that the quantities $z_s$ and $r^{\rm st}$ are
identical, so that we may rewrite Eq. (\ref{eq:saddle-point-equation}) as
\begin{equation}
r^{\rm st}=\frac{1}{M}\frac{\partial \ln \lambda_{\rm max}\left(\beta z,\frac{\beta
K}{M}\right)}{\partial(\beta z)}\Big|_{z=r^{\rm st}}.
\label{eq:saddle-point-equation-rst}
\end{equation}

In line with our set-out objective of obtaining the phase diagram, we
now need to solve Eq. (\ref{eq:saddle-point-equation-rst}) for $r^{\rm
st}$ as a function
of $\beta$ and $K$, in the limit $M \to \infty$. One has to then first compute the largest eigenvalue $\lambda_{\rm max}$
for finite $M$, then solve Eq. (\ref{eq:saddle-point-equation-rst}) for $r^{\rm
st}$, thereby obtaining the phase diagram in the $(K,T)$-plane for the
given value of $M$, and, finally, take the limit $M \to \infty$ of the
results. A roadblock in pursuing this program is the analytic computation of
$\lambda_{\rm max}$ for general $M$. One may alternatively estimate
$\lambda_{\rm max}$ numerically by discretizing the angles over the interval
$[0,2\pi)$, for example, as $\theta_j^{(a_j)}=a_j\Delta \theta$, with
$a_j=1,2,\ldots,P$ and $\Delta \theta=2\pi/P$ for any large positive
integer $P$; The operator ${\cal T}(C,C')$ then takes the form of a
matrix of size $P^M \times
P^M$, which even for any reasonable
value of $P$ and for not-so-large $M$ becomes numerically quite
unmanageable in order that we may reliably estimate using Eq.
(\ref{eq:saddle-point-equation-rst}) the phase diagram in the $(K,T)$-plane, leave alone the
limit that is of interest to us, namely, the limit $M \to \infty$. In order to gain insights into the phase transitions for finite $M$ and
their limiting behavior as $M \to \infty$, it proves insightful to consider
an equivalent Ising system for which the operator ${\cal T}(C,C')$ is
much more manageable numerically, and, consequently, the estimation of
$\lambda_{\rm max}$ is simpler and reliable, as we demonstrate in Section
\ref{sec:Ising-problem}. Our subsequent analysis and line of argument
will proceed along the following directions. We will first show for the
equivalent Ising system that in the
limit $M \to \infty$, the phase diagram in the $(K,T)$-plane obtained
numerically using the approach of the transfer operator may be
derived analytically by considering the model in the mean-field approximation. This observation hints at an apparent
mean-field dominance in dictating the stationary properties of the
equivalent Ising system in the limit $M \to \infty$. Assuming a similar
mean-field dominance to also be at work for the oscillator problem at hand,
Eq. (\ref{eq:eom-general}) with $\Delta=0$, we then perform an
explicit mean-field approximation of the model in equilibrium to
determine its phase diagram in the $(K,T)$-plane, and show that the
results are fully consistent with the phase diagram in the
$(\Delta,K)$-plane derived in Section \ref{sec:DeltaK-analysis}.

\subsubsection{An equivalent Ising problem}
\label{sec:Ising-problem}

To define an equivalent Ising problem, consider a setting similar to the one for the dynamics
(\ref{eq:eom-general}), namely, a $1d$ lattice of $N$ sites with periodic
boundary conditions, where
we take each site to be occupied by an Ising spin $S_j=\pm 1$. There is
an all-to-all ferromagnetic coupling between the spins. Additionally, each spin
interacts with strength $K/(4M)$ with $M$ neighboring spins to the left
and to the right, with $M<N/2$. The coupling $K$ can be of either sign,
with $K>0$ (respectively, $K<0$) implying a ferromagnetic (respectively,
an antiferromagnetic) $M$-neighbor interaction. The Hamiltonian of
the Ising system comprises just the potential energy ${\cal V}_{\rm
Ising}(\{S_j\})$ given by
\begin{equation}
{\cal V}_{\rm Ising}(\{S_j\})\equiv
\frac{1}{2N}\sum_{j,k=1}^N\left(1-S_jS_k\right)-\frac{K}{4M}\sum_{j=1}^N \sum_{k=-M}^M \left(S_j
S_{j+k}-1\right).
\label{eq:H-Ising}
\end{equation}
The two terms on the right hand side of Eq. (\ref{eq:H-Ising}) are just the
Ising analog of the corresponding terms on the right hand side of Eq. (\ref{eq:V}),
obtained by replacing the continuous variables $\theta_j$ in the latter
with the discrete Ising variables $S_j$. 
For $M=1$, the model (\ref{eq:H-Ising}) reduces to the Ising model with
nearest-neighbor 
and long-range interactions studied in Refs.
\cite{Nagle:1970,Bonner:1971,Kardar:1983,Mukamel:2005}. Moreover, in Eq.
(\ref{eq:H-Ising}), setting $K$ to zero allows to recover the mean-field
Ising model (the zero-field Curie-Weiss model of ferromagnet) \cite{Huang:1987,Salinas:2001}. The
magnetic order in the system is characterized by the magnetization
\begin{equation}
m\equiv \frac{\sum_{j=1}^N S_j}{N}.
\end{equation}

Similar to Eq. (\ref{eq:Delta0-equilibrium-distribution}), the Ising system has the BG
equilibrium distribution given by $P_{\rm eq}(\{S_j\}) \propto
\exp[-{\cal V}_{\rm Ising}(\{S_j\})/T]$, so that the canonical partition
function reads
\begin{eqnarray}
Z_N^{\rm Ising}=e^{-\beta N/2}\sum_{\{S_j=\pm 1\}}
\exp\left[
\frac{\beta}{2N}\Big(\sum_{j=1}^N S_j\Big)^2+\frac{\beta K}{4M}\sum_{j=1}^N\sum_{k=-M}^M \left(S_j
S_{j+k}-1\right)\right]. \nonumber \\
\label{eq:canonical-partition-function-Ising}
\end{eqnarray}
Invoking the Hubbard-Stratonovich transformation
(\ref{eq:hubbard-stratonovich-transformation}), and proceeding
similarly to the analysis presented in Sections \ref{sec:KT-analysis} and
\ref{sec:KT-analysis-Kneq0}, one obtains 
\begin{eqnarray}
Z_N^{\rm Ising}&=&\left(\frac{N\beta}{2\pi}\right)^{1/2}\int_{-\infty}^\infty 
{\rm d}z~\exp\left[-\frac{N\beta}{2}(1+z^2)\right]{\cal
Z}_N^{\rm Ising}; \label{eq:canonical-partition-function-auxiliary-field-Ising} \\
{\cal Z}_N^{\rm Ising}&\equiv&\sum_{\{S_j=\pm 1\}}
\exp\left[\beta z \sum_{j=1}^NS_j+\frac{\beta
K}{4M}\sum_{j=1}^N\sum_{k=-M}^M\left(S_jS_{j+k}-1\right)\right].
\label{eq:canonical-partition-function-Mneighbour}
\end{eqnarray}
The above equations are equivalents of Eqs.
(\ref{eq:canonical-partition-function-3}) and
(\ref{eq:canonical-partition-function-auxiliary-fields-nophi}). Similar to Eq.
(\ref{eq:canonical-partition-function-auxiliary-fields-nophi}), one may
interpret the factor ${\cal Z}_N^{\rm Ising}$ in Eq. (\ref{eq:canonical-partition-function-Mneighbour}) as the canonical partition function
of a $1d$ periodic chain of $N$ Ising spins, with each spin interacting with strength $K/(4M)$ with $M$
neighboring spins to the left and to
the right, and also with an external field of strength $z$. Employing
such an analogy, one may proceed to evaluate the factor ${\cal Z}_N^{\rm
Ising}$, by
following the same line of analysis involving the transfer operator that
was pursued in Section
\ref{sec:KT-analysis-Kneq0} to evaluate the factor ${\cal Z}_N$. A
difference that arises in the present
case of Ising spins
with respect to the oscillator case is that the transfer operator now takes the form of a $2^M \times 2^M$
matrix ${\cal T}^{\rm Ising}$, with elements given by
\begin{eqnarray}
&&{\cal T}^{\rm Ising}_{C,C'}\equiv
\exp\left(-\beta\Big[\frac{1}{2}X^{\rm Ising}_{C}+Y^{\rm
Ising}_{C,C'}+\frac{1}{2}X^{\rm Ising}_{C'}\Big]\right);
\label{eq:TCC-Ising} \\
&&X^{\rm Ising}_{C}\equiv -z\sum_{j=1}^M
S_{j}-\frac{K}{2M}\sum_{j=1}^M\sum_{k=1}^{M-j}\left(S_{j}S_{j+k}-1\right),
\label{eq:canonical-partition-function-XC-Ising} \\
&&Y^{\rm Ising}_{C,C'}\equiv-\frac{K}{2M}\sum_{j=1}^M
\sum_{k=1}^j\left(S^\prime_{j-k+1}S_{M+1-k}-1\right),
\label{eq:canonical-partition-function-YC-Ising} \\
&&C \equiv \{S_j\}_{1\le j \le M}. \label{eq:C-Ising}
\end{eqnarray}
Noting that ${\cal T}^{\rm Ising}_{C,C'}$ is a finite-dimensional real
square matrix
with positive entries, the application of the Perron-Frobenius theorem
\cite{Perron-Frobenius} implies the existence of its
largest eigenvalue $\lambda^{\rm Ising}_{\rm max}$ that is real and non-degenerate.
Hence, similar to Eqs.
(\ref{eq:canonical-partition-function-lambdamax-again}) and (\ref{eq:canonical-partition-function-saddle-point}), we obtain in the limit $N \to
\infty$ the result ${\cal Z}_N^{\rm Ising}=\left(\lambda^{\rm Ising}_{\rm
max}\right)^{N/M}$, and
consequently, 
\begin{equation}
Z_N^{\rm
Ising}=\left(\frac{N\beta}{2\pi}\right)^{1/2}\exp\left[-N\Big\{\frac{\beta}{2}(1+z_s^2)-\frac{1}{M}\ln
\lambda^{\rm Ising}_{\rm
max}\Big(\beta
z_s,\frac{\beta K}{M}\Big)\Big\}\right],
\label{eq:canonical-partition-function-saddle-point-Ising}
\end{equation}
where $z_s$ solves a saddle-point equation that has the same form as
in Eq. (\ref{eq:saddle-point-equation}). The dimenionsionless free
energy has the form of Eq. (\ref{eq:saddle-point-definition-2}).
Moreover, following the arguments given in Section
\ref{sec:KT-analysis-Kneq0} to show that the quantity $z_s$ in the
saddle-point equation (\ref{eq:saddle-point-equation}) is nothing but the stationary Kuramoto order parameter
$r^{\rm st}$, it may be shown that the quantity $z_s$ in Eq.
(\ref{eq:canonical-partition-function-saddle-point-Ising}) coincides
with the stationary magnetization $m^{\rm st}$.

\begin{figure}[!h]
\centering
\includegraphics[width=150mm]{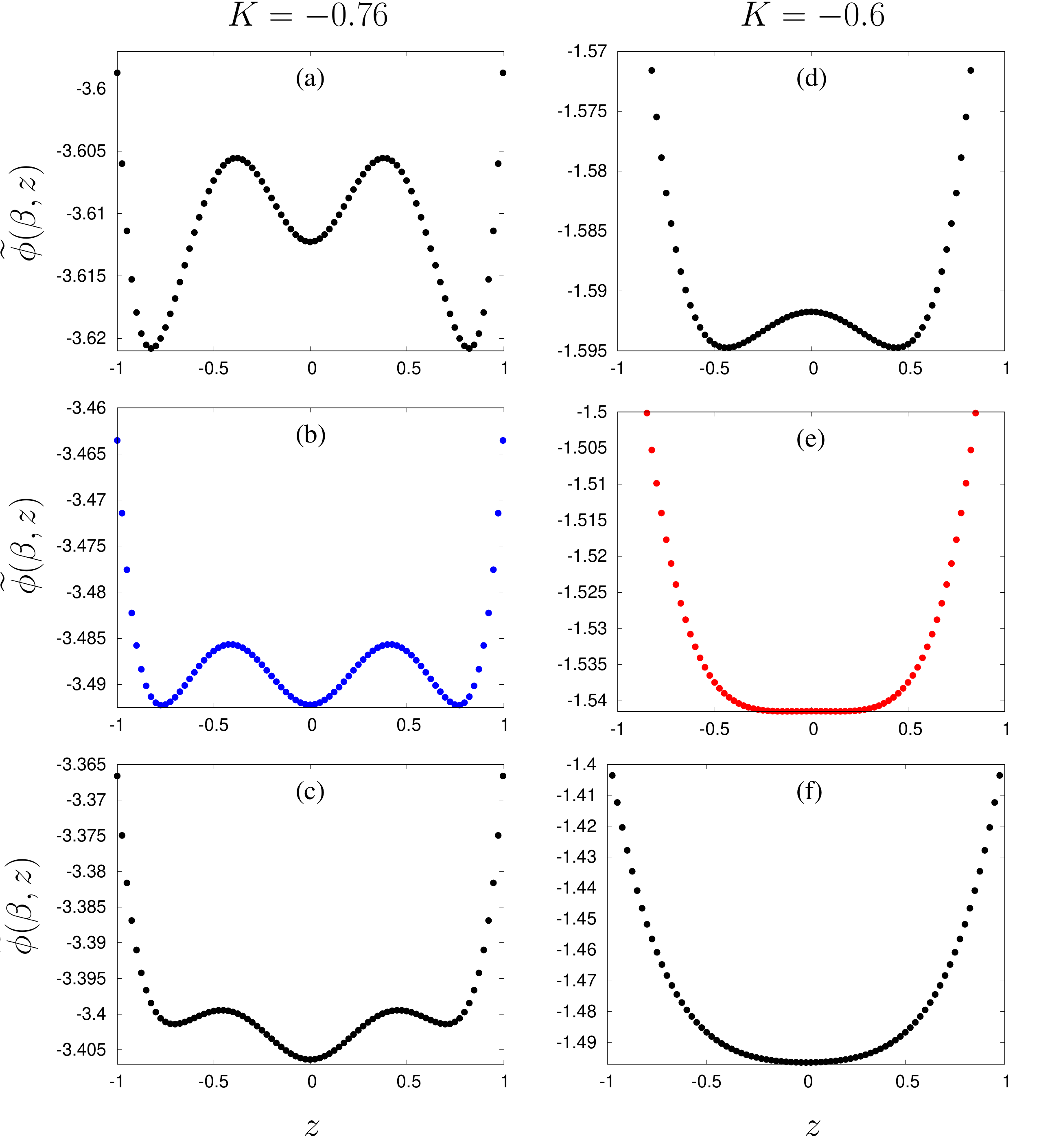}
\caption{For the mean-field Ising model with additional $M$-neighbor
interactions, (\ref{eq:H-Ising}), the figure shows the dimensionless
free-energy function as a function of $z$, Eq. (\ref{eq:saddle-point-definition-2}), for
$M=8$, and for three values of the temperature around a
phase transition. For $K=-0.76$, one has a first-order phase transition:
panel (b) refers to the transition temperature, while panel (a) (respectively,
panel (c)) refers to a temperature below (respectively, above) the
transition temperature. On the other hand, for $K=-0.6$, one has a
continuous phase transition: panel (e) refers to the transition
temperature,
while panel (d) (respectively,
panel (f)) refers to a temperature below (respectively, above) the
transition temperature. 
The points are obtained by estimating numerically the largest eigenvalue
$\lambda_{\rm max}$ of the transfer matrix (\ref{eq:TCC-Ising}), and
then using Eq. (\ref{eq:saddle-point-definition-2}).} 
\label{fig:Ising-free-energy}
\end{figure}

The basic program to identify the phase transition point $T_c(K,M)$ for
given values of $K$ and $M$ is as follows: 
\begin{itemize} 
\item For a given value of the temperature $T$, we first form the matrix
${\cal T}^{\rm Ising}$ in Eq.
(\ref{eq:TCC-Ising}), and then compute numerically its largest
eigenvalue $\lambda_{\rm max}$ by invoking the so-called power method
\cite{Larson:2017}; to this end, we employ a numerically efficient code that
implements the method \footnote{A FORTRAN90 library that implements the power
method and is distributed under the GNU LGPL license is available at
\url{http://people.sc.fsu.edu/~jburkardt/f_src/power_method/power_method.html}}.  
\item We then compute the free-energy function
$\widetilde{\phi}(\beta,z)$ as a function of $z$ by using Eq. (\ref{eq:saddle-point-definition-2}).  
\item
We repeat the last two steps for several values of $T$, locating
numerically for each $T$ the
value of $z$ at which $\widetilde{\phi}(\beta,z)$ is minimum. Because of
the symmetry of our problem, non-zero minimizers of
$\widetilde{\phi}(\beta,z)$, if and when they exist, always
occur in pairs symmetrically disposed on either side of zero: $z_s=\pm
A$, with $0<A<1$.  
\item A continuous phase transition point is given by the
value of $T$ at which the two non-zero
minimizers occurring at lower temperatures merge with each other for the
first time, so that the only minimizer at higher temperatures is at
$z_s=0$, see Fig. \ref{fig:Ising-free-energy}.
\item A
first-order phase transition point is given by the value of $T$ at which there
are three minimizers at $z_s=\pm A$ and at
$z_s=0$, with $0<A<1$, such that the values of
$\widetilde{\phi}(\beta,z)$ at these three minima coincide, see Fig.
\ref{fig:Ising-free-energy}.
\end{itemize}

\begin{figure}
\centering
\includegraphics[width=90mm]{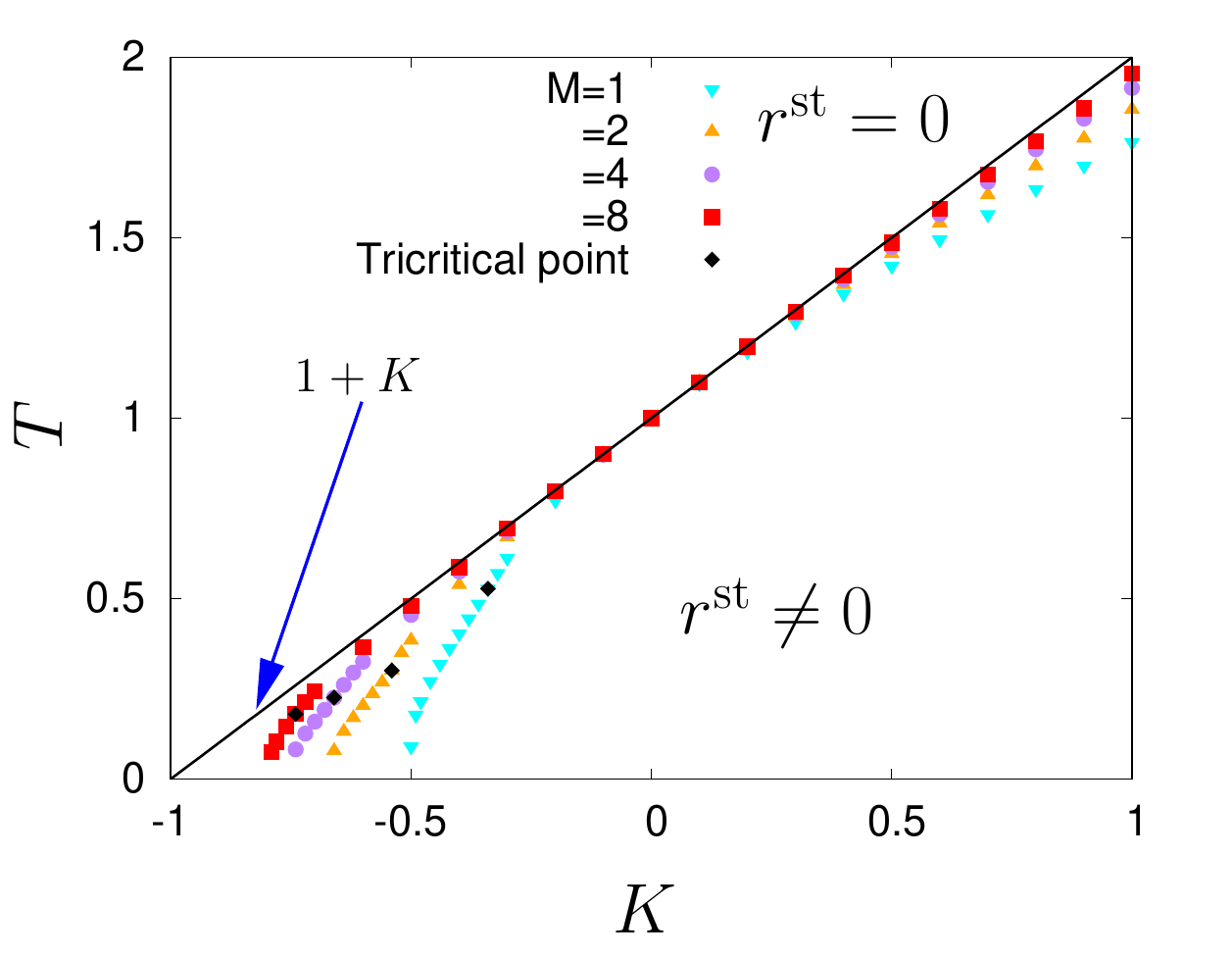}
\caption{For the mean-field Ising model with additional $M$-neighbor
interactions, (\ref{eq:H-Ising}), the figure shows the phase diagram in
the $(K,T)$-plane for four values of $M$: On increasing $T$ at a fixed $K$, the system
undergoes a phase transition from a synchronized phase ($r^{\rm st}\ne
0$) at low temperatures to an unsynchronized phase ($r^{\rm st}=0$) at
high temperatures. The phase transition is continuous for smaller (in
magnitude) values
of $K$, and is of first order for larger (in magnitude) values of $K$,
with the two separated by a tricritical point indicated in the figure.
The results are obtained by first estimating numerically the largest eigenvalue
$\lambda_{\rm max}$ of the transfer matrix (\ref{eq:TCC-Ising}), then
using Eq. (\ref{eq:saddle-point-definition-2}) to evaluate the free
energy, and finally studying at a fixed $K$ the
behavior of the minima of the free energy as a function of the
temperature. It may be observed from the figure that with increase of
$M$, the phase boundary approaches the line $T=1+K$, which is therefore
the $M\to \infty$ limit of the phase boundary.} 
\label{fig:Ising-phase-diagram}
\end{figure}

Following the above program, the phase diagram in the $(K,T)$-plane is
reported in Fig. \ref{fig:Ising-phase-diagram}. At a fixed $K$, as $T$
is increased, the system
undergoes a phase transition from a low-$T$ synchronized phase ($r^{\rm st}\ne
0$) to a high-$T$ unsynchronized phase ($r^{\rm st}=0$). The phase transition is continuous for smaller (in
magnitude) values
of $K$, and is of first order for larger (in magnitude) values of $K$,
with the two separated by a tricritical point indicated in the figure. For $M=1$, one
may solve exactly for the
line of continuous transition and the tricritical point, which are
respective given by $1/T=\exp\left(-K/T\right)$ and $K_{\rm
CTP}=-(\ln 3)/(2\sqrt{3})$ \cite{Nagle:1970,Bonner:1971}; Note that for
$K=0$, the equation $1/T=\exp\left(-K/(2T)\right)$ gives the phase
transition point of the mean-field Ising model as $T_c=1$
\cite{Salinas:2001}. From Fig. \ref{fig:Ising-phase-diagram}, it is evident that with increase of
$M$, (i) the tricritical point approaches the $K$-axis, and (ii) the
phase boundary approaches the line $T=1+K$, the latter being therefore
the $M\to \infty$ limit of the phase boundary. We may thus conclude that
in the limit $M\to \infty$, the model (\ref{eq:H-Ising}) exhibits a
continuous transition at all finite temperatures and a first-order
transition only at zero temperature, with the phase boundary given by
\begin{equation}
K_c^{\rm Ising}=T-1.
\label{eq:phase-boundary-KT-Ising}
\end{equation}
The above equation yields
the correct value for the phase transition point for $K=0$, and also gives a phase transition point at zero temperature for $K=-1$.

The result (\ref{eq:phase-boundary-KT-Ising}) may be understood
physically as follows. On utilizing the translational
invariance of the system, and on assuming pair
factorization $\langle S_j S_k \rangle_{\rm eq}=\langle S_j
\rangle_{\rm eq}\langle S_k \rangle_{\rm eq}$ in the joint limit $N \to
\infty, M \to \infty$ \footnote{The pair factorization is expected to hold exactly for
a purely mean-field model \cite{Huang:1987}.}, Eq. (\ref{eq:H-Ising}) gives the average energy density in
equilibrium as
\begin{equation}
\epsilon_{\rm Ising}=\frac{\left(1+K\right)}{2}\left(1-\langle m\rangle_{\rm
eq}^2\right).
\label{eq:H-Ising-1-eqavg-1}
\end{equation}
The above expression for the equilibrium energy density coincides (up to
an inconsequential constant term) with
that for the mean-field Ising model with coupling constant $J_{\rm
eff}^{\rm mean-field~Ising}\equiv
1+K$. At any non-zero temperature, the mean-field Ising model exhibits a continuous transition as a function of temperature, from a low-$T$ magnetized phase to a high-$T$
unmagnetized phase at the critical temperature $T_c^{\rm
mean-field~Ising}\equiv J_{\rm eff}^{\rm mean-field~Ising}=1+K$ \cite{Huang:1987,Salinas:2001}, from which one readily obtains Eq.
(\ref{eq:phase-boundary-KT-Ising}). At zero temperature, one has a
first-order transition between a magnetized phase in which all the spins
are aligned parallel to each other, and a non-magnetized phase in which
neighboring spins point in opposite directions.

\begin{figure}[!h]
\centering
\includegraphics[width=90mm]{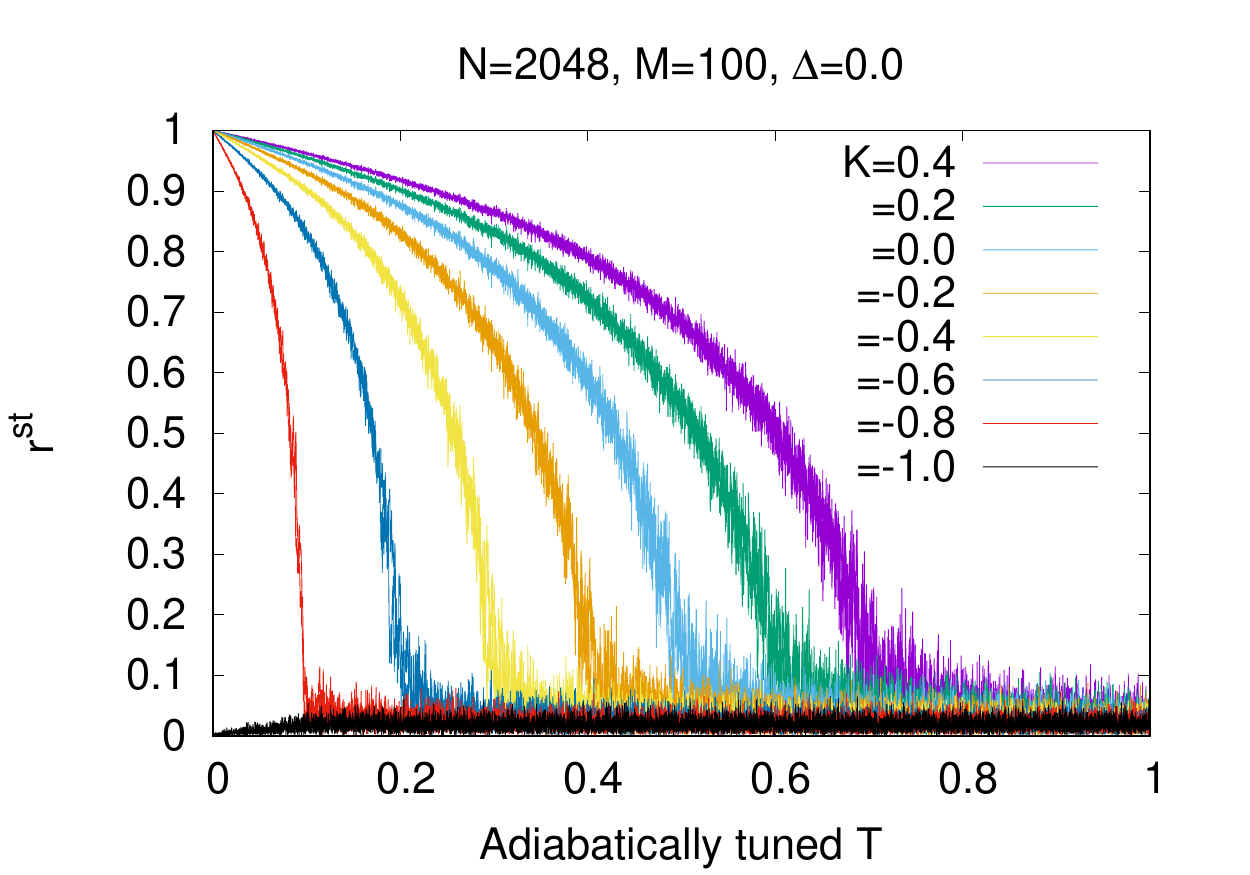}
\caption{For the Kuramoto model with additional $M$-neighbor
interactions, (\ref{eq:eom-general}), the figure
shows for $\Delta=0$ (the equilibrium limit) the variation in the synchronization order parameter $r^{\rm st}$ as a function of
adiabatically tuned $T$ for different values of the $M$-neighbor
coupling $K$. Starting with the stationary state at $T=0$, the
order parameter is monitored as $T$ 
is increased adiabatically as a function of time to high values and back in a cycle. The two
branches of each curve, corresponding to increasing and decreasing
values of $T$, almost overlap. We observe from the figure the absence of
hysteresis loops and abrupt jumps in the behavior of $r^{\rm st}$, which
would have hinted at the existence of a first-order transition.
Rather, the smooth variation of $r^{\rm st}$ with $T$ is
consistent with a continuous transition, and corroborates the
theoretical analysis of the main text. The data are obtained by numerical
integration of the dynamical equation (\ref{eq:eom-general}), by using the scheme detailed in
\ref{app:numerics-details}. The number of oscillators is $N=2048$, while
the value of $M$ used is $M=100$. We have checked that the results do not
change substantially for higher values of $M$.}
\label{fig:hysteresis-delta0}
\end{figure}

On the basis of the above analysis for the equivalent Ising problem,
(\ref{eq:H-Ising}), we may anticipate for the model
(\ref{eq:eom-general}) with $\Delta=0$ that in the limit $M \to \infty$,
the model exhibits a continuous transition at all temperatures $T >0$
and a first-order transition at $T=0$, with the phase boundary given by
\begin{equation}
K_c(T,\Delta=0)=2T-1.
\label{eq:phase-boundary-KT}
\end{equation}
Similar to the Ising case considered in the preceding paragraph, the
above equation may actually be derived by considering the equilibrium average of
Eq. (\ref{eq:V}), and by utilizing translational invariance and assuming pair factorization $\langle \sin \theta_j \sin
\theta_k\rangle_{\rm eq}=\langle \sin \theta_j \rangle_{\rm eq}\langle
\sin \theta_k \rangle_{\rm eq}$ and $\langle \cos \theta_j \cos
\theta_k\rangle_{\rm eq}=\langle \cos \theta_j \rangle_{\rm eq}\langle
\cos \theta_k \rangle_{\rm eq}$, to obtain the average energy density in
equilibrium as 
\begin{equation}
\epsilon=\frac{1}{2}-\frac{(1+K)}{2}(r^{\rm st})^2.
\label{eq:H-oscillators-eqavg}
\end{equation}
Here, we have used $(r^{\rm st})^2\equiv (r^{\rm eq})^2=\langle r_x \rangle_{\rm
eq}^2+\langle r_y \rangle_{\rm eq}^2$, as follows from Eq.
(\ref{eq:r-definition}). Up to an irrelevant constant term, Eq. (\ref{eq:H-oscillators-eqavg}) is the
same as the equilibrium energy density of the mean-field XY
model with effective coupling constant $J_{\rm eff}^{\rm mean-field~XY}\equiv1+K$, and
which exhibits a continuous phase transition at the critical temperature
$T_c^{\rm mean-field~XY}\equiv J_{\rm eff}^{\rm mean-field~XY}=
(1+K)/2$. Equation (\ref{eq:phase-boundary-KT}) yields consistently and correctly the phase transition point for $K=0$,
namely, $T_c(\Delta=0,K=0)=1/2$, the phase transition point of the BMF
model, see Fig. \ref{fig:schematic-phase-diagram}. The difference
between the Ising and the Kuramoto model in that the former deals with
discrete variables while the latter with continuous variables is
reflected in the appearance of an extra factor of two in Eq. (\ref{eq:phase-boundary-KT}) with respect to Eq.
(\ref{eq:phase-boundary-KT-Ising}). Let us note in passing that
considering an Ising ferromagnet with pair-wise interactions and the
classical XY model with the same couplings, the critical inverse
temperatures in the two cases have been proved to satisfy $\beta_c^{\rm XY}
\ge \beta_c^{\rm Ising}$ \cite{Aizenman:1980}; for the mean-field case of interaction that we
study here, we indeed find that the equality holds. 
We checked the result
(\ref{eq:phase-boundary-KT}) in direct simulations of the dynamics
(\ref{eq:eom-general}) by performing numerical integration of Eq.
(\ref{eq:eom-general}) with $\Delta=0$, by using the scheme detailed in
\ref{app:numerics-details}. The results, presented in Fig.
\ref{fig:hysteresis-delta0}, show the absence of
hysteresis loops and abrupt jumps characteristic of a first-order
transition, but rather a smooth variation of $r^{\rm st}$ with $T$
consistent with a continuous transition and in agreement with the
analysis in the foregoing paragraphs. 

%

\section{Simulation results for a general point in the
$(\Delta,K,T)$-space}
\label{sec:DeltaKT-analysis}

\begin{figure}
\centering
\includegraphics[width=90mm]{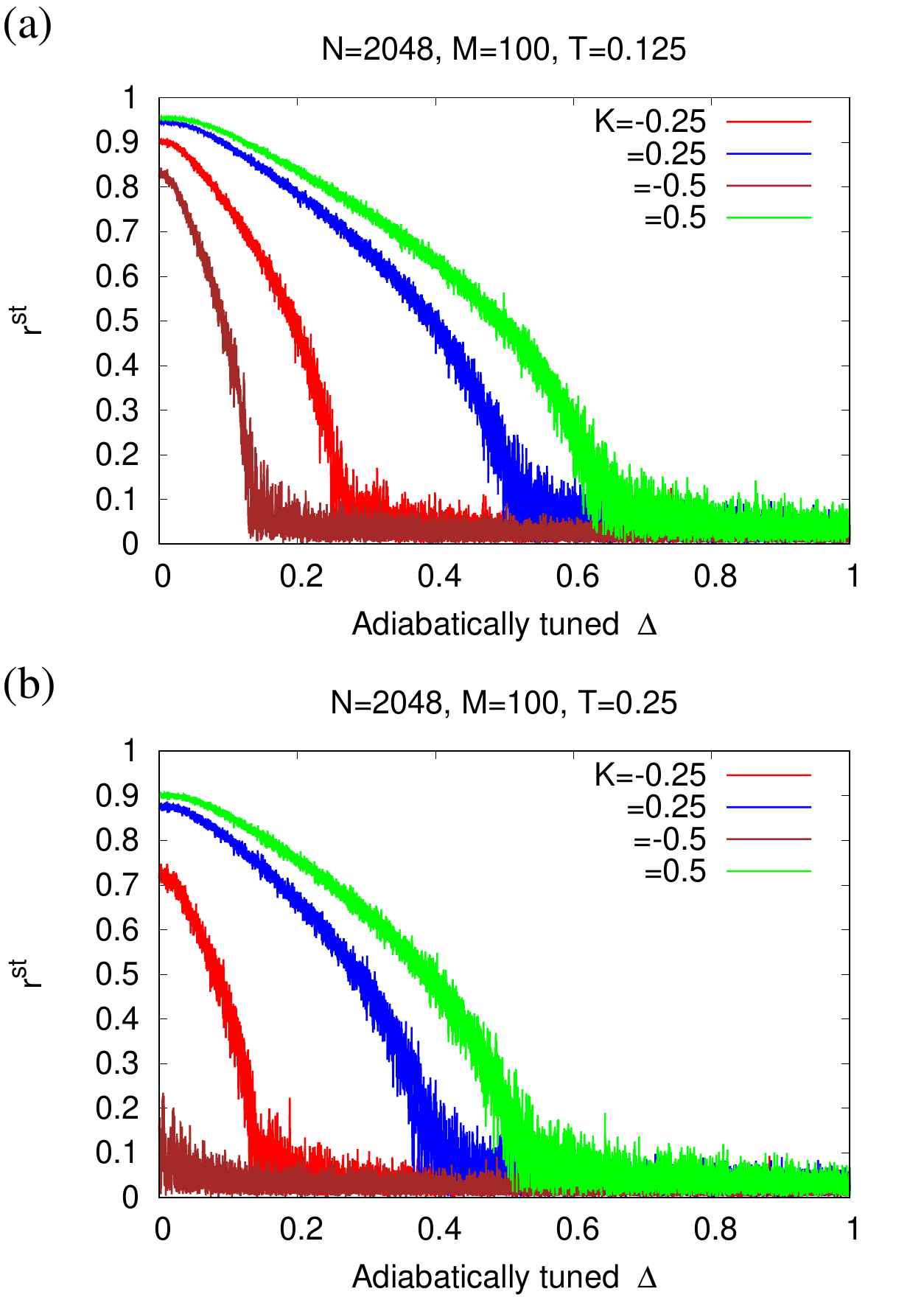}
\caption{Corresponding to the dynamics (\ref{eq:eom-general}) with
natural frequencies given by the Lorentzian distribution
(\ref{eq:lorentzian-g(w)}), the figure
shows the synchronization order parameter $r^{\rm st}$ as a function of
adiabatically tuned $\Delta$ for different values of the $M$-neighbor
coupling $K$ and two values of the temperature $T$. Starting with
equilibrium at $\Delta=0$, the order parameter is monitored as $\Delta$
is increased adiabatically as a function of time to high values and back in a cycle. The two
branches of each curve, corresponding to increasing and decreasing
values of $\Delta$, almost overlap. This is consistent with a continuous transition
and with the phase diagram in Fig. \ref{fig:schematic-phase-diagram}.
The data are obtained by numerical
integration of the dynamical equation (\ref{eq:eom-general}); for
details on the integration scheme, see \ref{app:numerics-details}. The number of oscillators is $N=2048$, while
the value of $M$ used is $M=100$. We have checked that the results do not
change substantially for higher values of $M$.}
\label{fig:hysteresis-general-point}
\end{figure}

In the absence of analytical results, in this section, we report on simulation
results on stationary state phase transitions for a general point in the
$(\Delta,K,T)$-space in Fig. \ref{fig:schematic-phase-diagram}. To
obtain the results, we performed numerical integration of Eq. (\ref{eq:eom-general}) for the Lorentzian
$g(\omega)$, Eq. (\ref{eq:lorentzian-g(w)}). For details on the
numerical scheme, see \ref{app:numerics-details}. For given values of $K$ and
$T$, and an initial configuration with oscillators at $\theta=0$, we let the
system equilibrate at $\Delta=0$. Subsequently, we tune $\Delta$
adiabatically to high values and back in a cycle. Note that the tuning of
$\Delta$ is performed for a fixed realization of the frequencies
$\omega_j$'s; Referring to Eq. (\ref{eq:eom-general}), we see that
tuning of $\Delta$ is equivalent to changing the factor multiplying the
frequency term in the equation of motion (\ref{eq:eom-general}). Adiabatic tuning of
$\Delta$ ensures that the system has sufficient time to attain stationarity before the value
of $\Delta$ changes significantly. Figure \ref{fig:hysteresis-general-point} shows the behavior of the synchronization order parameter $r^{\rm st}$
for several values of $K$ at representative temperatures and for a fixed
realization of the natural frequencies $\omega_j$'s. We have checked
that up to numerical precision, the results do not change on changing
the realization of the $\omega_j$'s. From the
figure, we observe the absence of sharp jumps and hysteresis behavior
characteristic of a first-order transition, but rather a continuous
variation of $r^{\rm st}$ with $\Delta$ expected of a continuous phase
transition. These features are consistent with the phase diagram shown
in Fig. \ref{fig:schematic-phase-diagram}, and lend credence to the analysis of the model
(\ref{eq:eom-general}) presented in this work. 

\begin{figure}
\centering
\includegraphics[width=90mm]{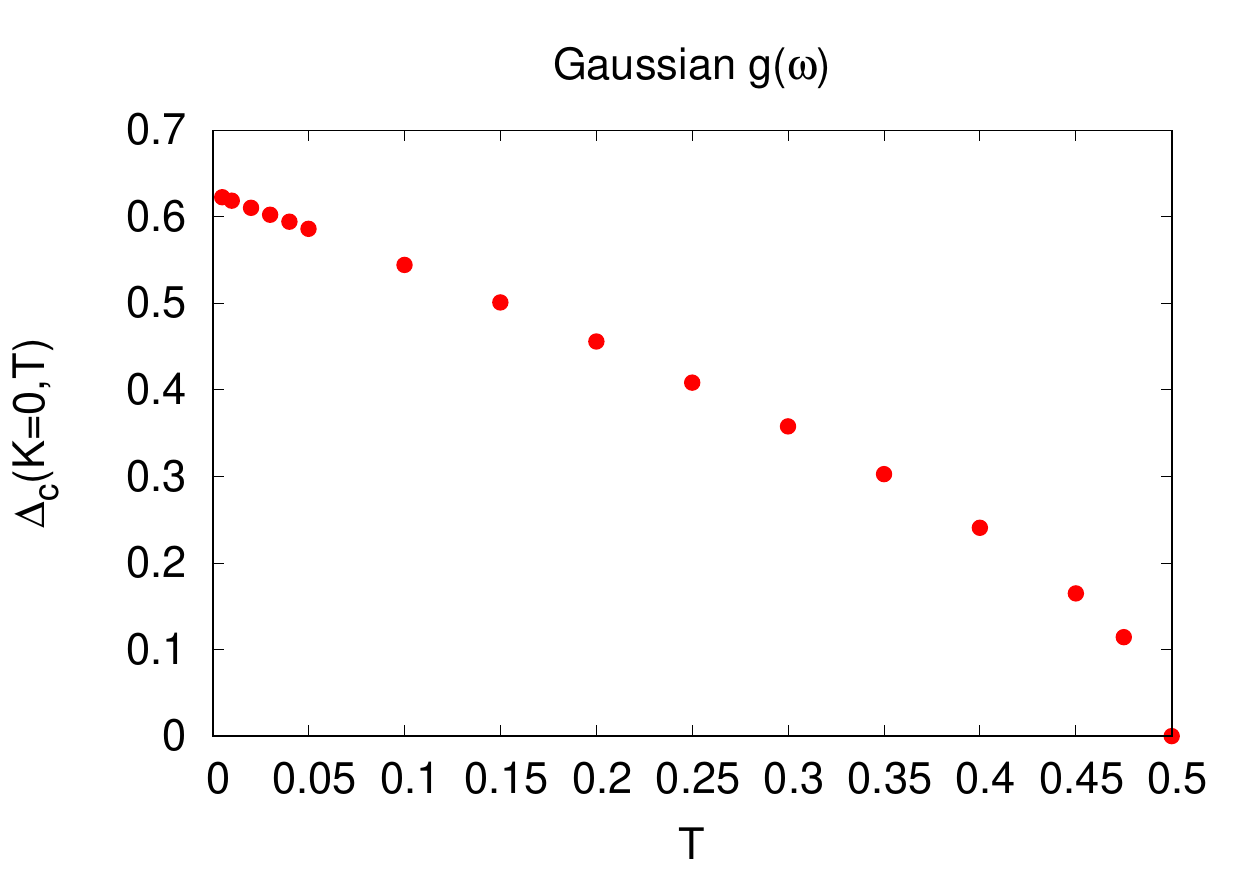}
\caption{Corresponding to the dynamics (\ref{eq:eom-general}) with
natural frequencies given by a Gaussian distribution
$g(\omega)=(1/\sqrt{2\pi})\exp(-\omega^2/2)$, the figure shows the
critical threshold $\Delta_c(K=0,T)$ obtained by solving numerically Eq.
(\ref{eq:Sakaguchi-transition-line}). At a fixed $T$, the system
undergoes a continuous transition from a low-$\Delta$ synchronized
($r^{\rm st}\ne 0$) phase
to high-$\Delta$ unsynchronized ($r^{\rm st}=0$) phase at the critical
threshold $\Delta_c(K=0,T)$.}
\label{fig:gaussian-Delta-T-transition line}
\end{figure}

\begin{figure}
\centering
\includegraphics[width=90mm]{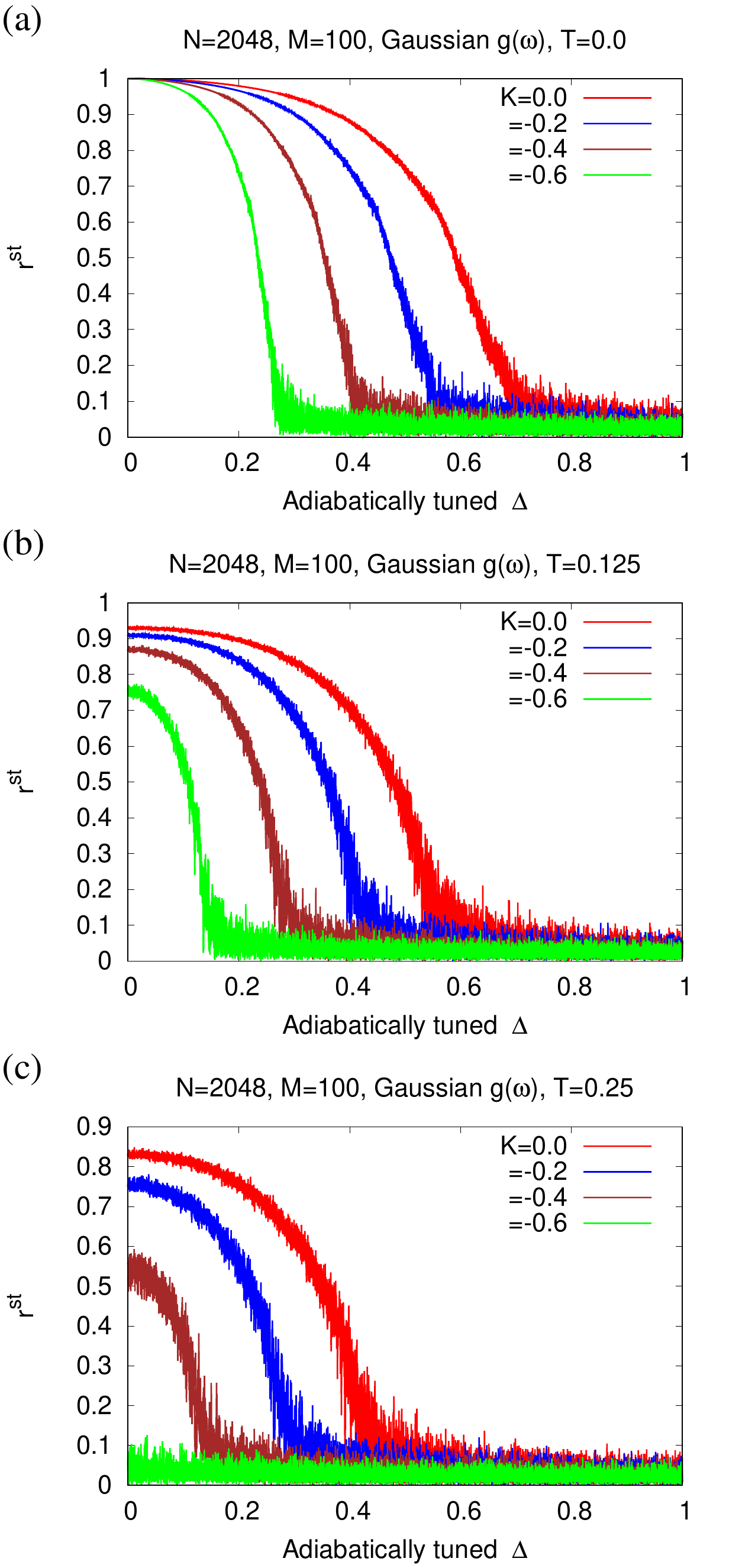}
\caption{Corresponding to the dynamics (\ref{eq:eom-general}) with
natural frequencies given by a Gaussian distribution
$g(\omega)=(1/\sqrt{2\pi})\exp(-\omega^2/2)$, the figure
shows the synchronization order parameter $r^{\rm st}$ as a function of
adiabatically tuned $\Delta$ for different values of the $M$-neighbor
coupling $K$ and the temperature $T$. Starting with
equilibrium at $\Delta=0$, the order parameter is monitored as $\Delta$
is increased adiabatically as a function of time to high values and back in a cycle. The two
branches of each curve, corresponding to increasing and decreasing
values of $\Delta$, almost overlap. This is consistent with a continuous transition
and with the phase diagram in Fig. \ref{fig:schematic-phase-diagram}.
The data are obtained by numerical
integration of the dynamical equation (\ref{eq:eom-general}); for
details on the integration scheme, see \ref{app:numerics-details}. The
number of oscillators is $N=2048$, while we have taken $M=100$. The results do not change
substantially for larger $M$.}
\label{fig:gaussian-hysteresis-general-point}
\end{figure}

Referring to the phase diagram \ref{fig:schematic-phase-diagram}, we
note that although the transition surface in the
$(\Delta,K,T)$-space intersects with the $(\Delta,K)$-plane, the
$(K,T)$-plane, and the $(\Delta,T)$-plane in straight lines, there is no
{\it a
priori} reason for the surface itself to be a plane. The simulation results in
Fig. \ref{fig:hysteresis-general-point} are however consistent with the equation
of a plane:
\begin{equation}
2\Delta+2T-K_c(T,\Delta)=1;
\label{eq:plane-DeltaKT}
\end{equation}
the above equation may be solved for $\Delta_c(K,T)$, and its value is
in good agreement with the transition point between the $r^{\rm
st} \ne 0$ and $r^{\rm st}=0$ phase observed in Fig.
\ref{fig:hysteresis-general-point}. In this regard, let us recall that
the phase diagram \ref{fig:schematic-phase-diagram} is obtained for a
Lorentzian distribution of the natural frequencies, Eq.
(\ref{eq:lorentzian-g(w)}). For a different
unimodal distribution with a non-compact support (e.g., a Gaussian), the intersection of the
transition surface with the $(K,T)$-plane, being applicable to the case
when the natural frequency term is absent in the dynamics, remains a
straight line, while its intersection with the $(\Delta,T)$-plane is
obtained from Eq. (\ref{eq:Sakaguchi-transition-line}). Let us choose a
Gaussian $g(\omega)$ given by
$g(\omega)=(1/\sqrt{2\pi})\exp(-\omega^2/2)$, for which the latter intersection is shown in
Fig. \ref{fig:gaussian-Delta-T-transition line}, and is evidently not a
straight line. The simulation
results for a Gaussian $g(\omega)$ are shown in Fig.
\ref{fig:gaussian-hysteresis-general-point}. Similar to the Lorentzian
case, one finds the absence of sharp jumps and hysteresis behavior
characteristic of a first-order transition, but rather a continuous
variation of $r^{\rm st}$ with $\Delta$ that implies a continuous phase
transition. From the figure, one may estimate the transition point between the $r^{\rm
st} \ne 0$ and $r^{\rm st}=0$ phase, and find from the estimated values
that the general transition surface in the $(\Delta,K,T)$-space for a
Gaussian frequency distribution is not a plane (and thus its intersection with the
$(\Delta,K)$-plane is not a straight line). Thus, we are led to conclude
that having straight transition lines in the $(\Delta,T)$-plane and the
$(\Delta,K)$ plane in Fig. \ref{fig:schematic-phase-diagram} is typical
to a Lorentzian frequency distribution and does not hold in general for
other unimodal distributions with a non-compact support. 

We observe a peculiar feature of the phase diagram \ref{fig:schematic-phase-diagram}: the
phase transition line in the $(K,T)$-plane and the
$(\Delta,K)$-plane suggests that the temperature $T$ in the dynamics
corresponding to the former plane plays a role similar to $\Delta$ in
the dynamics for the latter plane. Indeed, the
transition line in the $(K,T)$-plane is
$K_c(T,\Delta=0)=2T-1$, while the one in the $(\Delta,K)$-plane is
$K_c(T=0,\Delta)=2\Delta-1$. This observation is somewhat counterintuitive, given
that $\Delta$ corresponds to a quenched disordered (that is, a {\it
time-independent}) noise in the dynamics, while $T$ signifies an
annealed (that is, a {\it time-dependent}) noise, and that these
two types of noise typically have very different consequences on the properties of
many-body interacting systems, e.g., on surface growth dynamics
\cite{Meakin:1998}.

Another point worth noting about the
phase diagram \ref{fig:schematic-phase-diagram} is that both the lines
$K_c(T,\Delta=0)=2T-1$ and $K_c(T=0,\Delta)=2\Delta-1$ may be derived by
considering the noisy Kuramoto model with an
effective global coupling equal to $1+K$ (and thus only mean-field and
no non-local interaction), as we demonstrate below. To
this end, consider the 
equation of motion 
\begin{equation}
\frac{{\rm d}\theta_j}{{\rm d}t}=\Delta~\omega_j+\frac{(1+K)}{N}\sum_{k=1}^N
\sin(\theta_k-\theta_j)+\eta_j(t). 
\label{eq:eom-modified-global-coupling}
\end{equation}
Then, in the $(\Delta,K)$-plane, the model reduces to the Kuramoto model
with modified global coupling given by $J_{\rm eff}^{\rm Kuramoto}
\equiv 1+K$. In terms of rescaled variable $\widetilde{t}\equiv t(1+K)$,
the equation of motion has the form of Eq. (\ref{eq:eom-kuramoto-again})
for the Kuramoto model:
\begin{equation}
\frac{{\rm d}\theta_j}{{\rm d}\widetilde{t}}=\Delta_{\rm eff}~\omega_j+\frac{1}{N}\sum_{k=1}^N
\sin(\theta_k-\theta_j),
\label{eq:eom-modified-global-coupling-again}
\end{equation}
with
\begin{equation}
\Delta_{\rm eff}\equiv \frac{\Delta}{1+K}.
\label{eq:Delta-effective}
\end{equation}
Using the known results about the phase transition in the Kuramoto
model, we conclude that the dynamics (\ref{eq:eom-modified-global-coupling-again}) exhibits a
phase transition between a low-$\Delta_{\rm eff}$ synchronized phase
and a high-$\Delta_{\rm eff}$ unsynchronized phase at a critical value
given for the Lorentzian distribution (\ref{eq:lorentzian-g(w)}) by
$\left[\Delta_{\rm eff}\right]_c=\pi g(0)/2=1/2$. Combining this result
with Eq.
(\ref{eq:Delta-effective}), we obtain the result we had set out to
demonstrate, namely, $K_c(T=0,\Delta)=2\Delta-1$. To obtain the other
result, namely, $K_c(T,\Delta=0)=2T-1$, we start with Eq.
(\ref{eq:eom-modified-global-coupling}) with $\Delta=0$. Noting that
in this case, the dynamics relaxes to a 
Boltzmann-Gibbs (BG) equilibrium state with the probability distribution
of the angles $P_{\rm eq}(\{\theta_j\})\propto \exp[-\widetilde{\cal
V}(\{\theta_j\})/T];~\widetilde{\cal V}(\{\theta_j\})\equiv
(1+K)/(2N)\sum_{j,k=1}^N\left[1-\cos(\theta_j-\theta_k)\right]$, one
may perform an analysis similar to that in Section \ref{sec:KT-analysis-K0}
to arrive at the result $T_c(\Delta=0,K)=(1+K)/2$, which then yields
$K_c(T,\Delta=0)=2T-1$. 
The above derivation of the results $K_c(T=0,\Delta)=2\Delta-1$ and
$K_c(T,\Delta=0)=2T-1$ based on only mean-field interaction points towards an
apparent {\it mean-field dominance} in the stationary state of the dynamics
(\ref{eq:eom-general}) in the $(K,T)$-plane, where one has an
equilibrium dynamics, and in the $(\Delta,K)$-plane for a Lorentzian distribution, where one has a
non-equilibrium dynamics. Mean-field dominance in the stationary
state due to an equilibrium
\cite{Gupta:2012-1} and a nonequilibrium \cite{Gupta:2012} dynamics has
been recently observed in a variant of the Kuramoto model that comprises oscillators interacting with one another with a
strength that decays as a power-law of their separation on a $1d$
lattice \cite{Anteneodo:1998}. The origin of the
features of the phase diagram mentioned in the present and the two preceding
paragraphs and the necessary and
sufficient conditions for their validity are open issues left for future studies.

%

\section{Conclusions and perspectives}
\label{sec:conclusions}
In this work, we addressed the issue of spontaneous collective
synchronization in many-body interacting systems within the ambit of the
paradigmatic Kuramoto model of globally-coupled
phase oscillators of distributed natural frequencies. The model is
known to exhibit as a function of the diversity of the
natural frequencies a transition between a synchronized and an
unsynchronized phase. Our objective in this work was to investigate the
robustness of such a behavior with respect to additional interactions.
Specifically, we considered the effect of including a non-local $M$-neighbor interaction between the oscillators residing on the
sites of a one-dimensional
periodic lattice of $N$ sites. Here, we dealt with the case of a unimodal
frequency distribution, and in particular, a Lorentzian distribution. In presence of thermal noise of strength proportional
to a temperature, the resulting dynamics is effectively characterized by
three parameters: the width $\Delta$ of the frequency distribution, the
strength $K$ of the $M$-neighbor interaction, and the temperature $T$. 
In obtaining our results, we considered the simultaneous limits $M \to \infty, N \to \infty$,
while keeping the interaction radius $\sigma \equiv M/N$ to satisfy
$\sigma < 1/2$. The latter condition allows to have distinct forms of mean-field and
non-mean-field interactions.

The analysis presented in this work revealed that in the stationary
state, the dynamics of our model in the $(\Delta,K,T)$-space exhibits a very rich
phase diagram that involves both equilibrium and non-equilibrium phase transitions.
In two contrasting limits of the dynamics, namely, (i) the limit $T \to
0$, and (ii) the limit $\Delta \to 0$, we obtained exact analytical
results for the phase transitions, by borrowing tools from diverse
disciplines, namely, the ones of non-linear dynamics and statistical
mechanics. For case (i), when the dynamics reduces to that of a
non-linear {\it dynamical} system, our
exact analysis is based on the use of the so-called
Ott-Antonsen (OA) ansatz to derive a reduced set of nonlinear partial
differential equations for the macroscopic evolution of the system. 
On the other hand, in the case of (ii), the
dynamics becomes that of a {\it statistical} system in contact with a heat bath
and relaxing to a statistical equilibrium state, and our analytical
results are derived by extending the transfer matrix approach
of the nearest-neighbor Ising model to consider non-local
interactions. Referring to Fig. \ref{fig:schematic-phase-diagram}, the
line $K_c(T=0,\Delta)$ is obtained by the OA ansatz, while the line
$K_c(T,\Delta=0)$ is obtained by the transfer matrix method. Being an
ansatz, it is remarkable that the OA approach is able to correctly
predict the phase transition point $K_c(T=0,\Delta=0)=-1$, as is checked
by obtaining the same point from the analysis for the $(K,T)$-plane by employing the well-established and exact transfer matrix
approach \cite{Huang:1987}. It remains an outstanding problem to obtain analytical
results for the phase transition at a general point in the
$(\Delta,K,T)$-space.

While the analytical results presented using the Ott-Antonsen ansatz
applies to a Lorentzian distribution of the natural frequencies,
qualitatively similar results are expected to hold for other unimodal
frequency distributions that have a non-compact support similar to the
Lorentzian distribution. Referring to the phase diagram
\ref{fig:schematic-phase-diagram}, the phase boundary in the
$(K,T)$-plane is obviously independent of the choice of the frequency
distribution, while the boundaries in the $(\Delta,K)$- and the
$(\Delta,T)$-plane and, more generally, the transition surface in the
$(\Delta,K,T)$-space would depend on the specific form of the frequency
distribution. Nevertheless, we expect the general features of the phase
diagram to hold for other non-compact unimodal distributions, but these would certainly
change if one considers distributions that are unimodal with compact
support or those that are not unimodal, e.g., a
bimodal distribution \cite{Montbrio:2006,Pazo:2009,Martens:2009}. Resolution of this issue is under
investigation. Another open issue is to consider finite values of $M$.
In this case, the phase diagram in the $(K,T)$-plane shows a
tricritical point \cite{Dauxois:2010}, similar to the one observed for
the Ising case in Fig. \ref{fig:Ising-phase-diagram}, and we may expect the tricritical point to extend to a
tricritical line in the $(\Delta,K,T)$-space. Our preliminary results
indeed point in that direction, and a detailed analysis will be reported
elsewhere \cite{Gupta:2017-1}. 

We may mention other immediate and
physically relevant offshoots of our work, for example, considering in
the dynamics (\ref{eq:eom}) the global coupling term to also include a
second harmonic $\sim \sin\left(2(\theta_k-\theta_j)\right)$
\cite{Hansel:1993,Clusella:2016}, the presence of a local potential \cite{Campa:2016}, a phase-lag parameter
\cite{Sakaguchi:1986,Wolfrum:2011}, a
time delay in the interaction between the oscillators
\cite{Ott:2008,Laing:2016}, and/or 
considering in place of the first-order dynamics investigated in this work
the case of a second-order dynamics that includes the effect of a finite
inertia of the oscillators, and which is known to alter significantly
the behavior of the bare Kuramoto model
\cite{Gupta:2014-1,Komarov:2016,Campa:2015,Olmi:2015,Barre:2016}. 

\section{Acknowledgments}
\label{sec:acknowledgements}
This paper is dedicated to my beloved father, the most important
person in my life. SG is grateful to Alessandro Campa for fruitful discussions and
suggestions, and for a careful reading of the manuscript. SG is also grateful to the Laboratoire de Physique, \'{E}cole Normale
Sup\'{e}rieure de Lyon for support and warm hospitality during
his stay as Professeur Invit\'{e} in June 2017 when this manuscript was
being finalized.

\appendix

\section{\\ Proof that the dynamics (\ref{eq:eom-general}) does not satisfy detailed balance
unless $\Delta=0$}
\label{app:detailed-balance-proof}

In this Appendix, we give a formal proof that the dynamics
(\ref{eq:eom-general}) does not satisfy detailed balance in the
stationary state unless one has $\Delta=0$. To this end, we first consider for simplicity of
discussion the case of a
bimodal $g(\omega)$, and then generalize our discussion to a
general $g(\omega)$. Consider a given realization of $g(\omega)$, in which there are $N_1$
oscillators with natural frequencies equal to $\omega_1$ and $N_2$
oscillators with frequencies equal to $\omega_2$, with $N_1+N_2=N$. Note that we need to consider at least two values of the
natural frequencies in order to have a non-zero $\Delta$. Let us define the $N$-oscillator distribution function
$f_N(\theta_1,\dots,\theta_{N_1},\theta_{N_1+1},\dots,\theta_N,t)$
as the probability density at time $t$ to observe the system around the
values $\{\theta_j\}_{1\le j\le N}$, with the
normalization $\int\left(\prod_{j=1}^{N}{\rm d}
\theta_j\right)f_{N}(\{\theta_j\},t)=1$. 
The time evolution of $f_N$ follows the $N$-dimensional Fokker-Planck
equation that may be written down from the equation of motion
(\ref{eq:eom-general}) by following standard prescription
\cite{Gardiner:1983}: 
\begin{eqnarray}
&&\frac{\partial f_{N}}{\partial t}
=-\Delta\sum_{j=1}^{N}\left(\Omega^{T}\right)_j\frac{\partial f_N}{\partial
\theta_j}+T\sum_{j=1}^{N}\frac{\partial^{2}f_{N}}{\partial
\theta_j^{2}}\nonumber \\
&&-\frac{1}{N}\sum_{j,k=1}^{N}\frac{\partial }{\partial
\theta_j}\left(f_N\sin(\theta_k-\theta_j)\right)
\nonumber \\
&&-\frac{K}{2M}\sum_{j=1}^{N}\sum_{k=-M}^M\frac{\partial }{\partial
\theta_j}\left(f_N\sin(\theta_{j+k}-\theta_j)\right), 
\label{eq:fp-eqn}
\end{eqnarray}
where we have defined the $N\times 1$ column vector $\Omega$ with first
$N_1$ entries equal to $\omega_1$ and the following $N_{2}$ entries
equal to $\omega_2$, and where the superscript $T$ denotes matrix transpose
operation: $\Omega^T\equiv\left[\omega_{1} ~\omega_{1}
\dots~\omega_{1}~\omega_{2}\dots ~\omega_{2}\right]$.

The Fokker-Planck equation (\ref{eq:fp-eqn}) may be rewritten as
\begin{eqnarray}
&&\frac{\partial f_N({\bf x})}{\partial
t}=-\sum_{j=1}^N\frac{\partial(A_j({\bf x})f_N({\bf x}))}{\partial
x_{j}}+\frac{1}{2}\sum_{j,k=1}^{N}\frac{\partial^2(B_{j,k}({\bf
x})f_N({\bf x}))}{\partial x_j\partial x_k},
\label{eq:fp-compact}
\end{eqnarray}
where we have defined 
\begin{eqnarray}
&&x_j \equiv \theta_j;~j=1,2,\dots,N, \nonumber \\
\label{eq:x-defn} \\
&&{\bf x}=\{x_j\}_{1\le j \le N}. \nonumber
\end{eqnarray}
In Eq. (\ref{eq:fp-compact}), the drift vector $A_j({\bf x})$
is given by
\begin{equation}
A_j({\bf x})
\equiv\frac{1}{N}\sum_{k=1}^{N}\sin(\theta_k-\theta_j)+\frac{K}{2M}\sum_{k=-M}^M\sin(\theta_{j+k}-\theta_j)+\Delta~(\Omega^{T})_j,
\label{eq:Aj-defn}
\end{equation}
while the diffusion matrix is 
\begin{equation}
B_{j,k}({\bf x}) \equiv 2T\delta_{jk}.
\label{eq:Bjk-defn}
\end{equation}

The dynamics described by the Fokker-Planck equation of the form (\ref{eq:fp-compact})
satisfies detailed balance if and only if the following conditions
are satisfied \cite{Gardiner:1983}:
\begin{eqnarray}
&&\epsilon_j\epsilon_k B_{j,k}(\epsilon{\bf x}) =B_{j,k}({\bf x}),\label{eq:detailed-balance-cond1}\\
&&\epsilon_j A_j(\epsilon{\bf x})f_{N}^{s}({\bf x})
=-A_j({\bf x})f_{N}^{s}({\bf x})+\sum_{k=1}^{N}\frac{\partial
B_{j,k}({\bf x})f_{N}^{s}({\bf x})}{\partial x_k},
\label{eq:detailed-balance-cond2}
\end{eqnarray}
where $f_{N}^{s}({\bf x})$ is the stationary solution of Eq. (\ref{eq:fp-compact}).
Here, $\epsilon_j \equiv \pm1$ denotes the parity of $x_j$'s with
respect to time reversal $t \to -t$: Under time reversal, the $x_j$'s transform as $x_j \rightarrow \epsilon_j x_j$,
with $\epsilon_j=-1$ or $+1$ depending on whether $x_j$ is
odd or even under time reversal. In our case, $\theta_j$'s
are even variables, so that we consider $\epsilon_j=+1~\forall~j$ in the following
discussion.

Using Eq. (\ref{eq:Bjk-defn}), we see that the condition (\ref{eq:detailed-balance-cond1}) is trivially satisfied for our model. To check
the other condition, namely, Eq. (\ref{eq:detailed-balance-cond2}), let us formally
solve the equation to obtain $f_N^s({\bf x})$, and check whether the
solution solves Eq. (\ref{eq:fp-compact}) in the stationary state. Using
$\epsilon_j=+1~\forall~j$, we see that the condition reduces to 
\begin{eqnarray}
A_j({\bf x})f_{N}^{s}({\bf x}) &
=-A_j({\bf x})f_{N}^{s}({\bf x})+2T\frac{\partial
f_{N}^{s}({\bf x})}{\partial \theta_j},
\label{eq:detailed-balance-cond2-again}
\end{eqnarray}
solving which yields 
\begin{eqnarray}
&&f_{N}^{s}({\bf x})\propto
\exp\Big[\frac{1}{T}\Big(\frac{1}{N}\sum_{j,k=1}^{N}\cos(\theta_k-\theta_j)\nonumber
\\
&&+\frac{K}{4M}\sum_{j=1}^N\sum_{k=-M}^M\cos(\theta_{j+k}-\theta_j)+\Delta\sum_{j=1}^N(\Omega^{T})_j\theta_j\Big)\Big].
\label{eq:stationary-soln1}
\end{eqnarray}
Substituting Eq. (\ref{eq:stationary-soln1}) into Eq. (\ref{eq:fp-compact}),
and requiring that $f_N^s({\bf x})$ given by the former is a stationary
solution of the latter, we obtain the condition that $\Delta$
has to be equal to zero. We thus conclude from the foregoing discussions that the dynamics
(\ref{eq:eom-general}) does not satisfy detailed balance unless
$\Delta=0$. The foregoing discussions for a bimodal $g(\omega)$,
establishing the lack of detailed balance for $\Delta \ne 0$, obviously
extend to any choice of $g(\omega)$. For $\Delta=0$, we get the stationary
solution as
\begin{equation}
f_{N,\Delta=0}^{s}({\bf x})\propto
\exp[-{\cal V}(\{\theta_j\})/T],
\label{eq:bmf-soln}
\end{equation}
where the function ${\cal V}(\{\theta_j\})$ is given by Eq.
(\ref{eq:V}). We thus obtain for $\Delta=0$ the equilibrium
distribution for the angles as
\begin{equation}
P_{\rm eq}(\{\theta_j\}) \propto
\exp[-{\cal V}(\{\theta_j\})/T].
\label{eq:bmf-soln-1}
\end{equation}
%
\section{\\ Numerical scheme to integrate the equation of motion
(\ref{eq:eom-general})}
\label{app:numerics-details}

Here we give details of the numerical scheme to integrate the equation of motion
(\ref{eq:eom-general}). To this end, rewriting the mean-field term in
Eq. (\ref{eq:eom-general}) in terms of the quantities $(r_x,r_y)$
defined in the paragraph following Eq. (\ref{eq:r-definition}), we obtain the net torque
acting on the $j$-th oscillator as
\begin{eqnarray}
F_j(t) &\equiv& \Delta~\omega_j+r_y(t) \cos \theta_j(t)-r_x(t) \sin
\theta_j(t)\nonumber \\
&&+\frac{K}{2M}\sum_{k=-M}^M
\cos\left(\theta_{j+k}(t)-\theta_j(t)\right).
\label{eq:Fj-definition}
\end{eqnarray}
To simulate the dynamics (\ref{eq:eom-general}) over a time interval
$[0:{\cal T}]$ and for given values of $K,M$ and $T$, we first
choose a time step size $\Delta t \ll 1$, and set $t_n=n\Delta t$ as the
$n$-th time step of the dynamics, where $n = 0,1,2, \ldots,{\cal N}_t$,
and ${\cal N}_t={\cal T}/\Delta t$. One step of the numerical scheme
then involves the following update of the $\theta_j$'s for $j=1,2,\ldots,N$:
\begin{eqnarray}
&&\theta_j\left(t_n+\Delta t\right)\nonumber \\
&&=\left\{
\begin{array}{ll}
               \theta_j(t_n)+\left(F_j(t_n)\Delta
t+\dot{F}(t_n)\frac{(\Delta t)^2}{2}\right)+X_n(\Delta t) &
               \mbox{for $n \ge 1$}, \nonumber \\
                \theta_j(t_n)+F_j(t_n)\Delta
t+X_n(\Delta t) & \mbox{if $n=0$},
               \end{array}
        \right. \\
\label{eq:thetaj-update}
\end{eqnarray}
with $X_n$ a Gaussian-distributed random number with zero mean and
variance given by
\begin{equation}
\langle X_n^2(\Delta t)\rangle=2T\Delta t,
\label{eq:Xn-variance}
\end{equation}
and
\begin{equation}
\dot{F}(t_n) \equiv \frac{F(t_n)-F(t_{n-1})}{\Delta t}.
\label{eq:Fdot-definition}
\end{equation}
The above numerical scheme neglects terms of order
higher than $\Delta t$, and may be derived by following the general
techniques discussed in Ref. \cite{Kloeden:1999}. For the numerical results reported in this work, we have chosen a fixed time step $\Delta
t=0.01$. We have checked that up to numerical precision, the obtained
numerical results do not depend on $\Delta t$, so long as the latter is
small.


\end{document}